\documentclass[useAMS]{mn2e}
\usepackage{times}
\usepackage{color}
\usepackage{epsf}
\definecolor{grey}{rgb}{0.5,0.6,0.7}
\definecolor{orange}{rgb}{0.8,0.7,0}

\def\gsim{\,\lower2truept\hbox{${>\atop\hbox{\raise4truept\hbox{$\sim$}}}$}\,}
\def\lsim{\,\lower2truept\hbox{${<\atop\hbox{\raise4truept\hbox{$\sim$}}}$}\,}
\def\msol{{\rm M}_\odot}
\def\mpc{{\rm Mpc}}
\def\kpc{{\rm kpc}}
\def\mdelta{\rm M_\Delta}
\def\rdelta{\rm r_\Delta}

\def\rvir{\rm r_{vir}}
\def\rdec{\rm r_{\rm dec}}
\def\nhr{n_{\rm HR}}
\def\nlr{n_{\rm LR}}
\def\vcirc{V_{\rm c}}
\def\magb{{\rm M}_{\rm B}}
\def\magv{{\rm M}_{\rm V}}
\def\magk{{\rm M}_{\rm K}}
\def\gal{GalICS}
\def\galone{{\sc galics~i}}
\def\mstar{{\rm M_*}}

\title{GALICS -- VI. Modelling Hierarchical Galaxy Formation in Clusters}
\author[Barbara Lanzoni et al.]
{B.~Lanzoni,$^1$\thanks{barbara.lanzoni@bo.astro.it} B.~Guiderdoni,$^2$
G.A.~Mamon,$^{3,4}$ J.~Devriendt,$^2$ S.~Hatton$^3$ \\ 
$^1$INAF-Osservatorio Astronomico di Bologna, Via Ranzani 1, 40125 Bologna,
Italy\\
$^2$ Observatoire Astronomique de Lyon, 9 avenue Charles Andr\'e, 69561 
Saint-Genis-Laval Cedex\\ 
$^3$Institut d'Astrophysique de Paris (CNRS UMR 7095), 98 bis Boulevard Arago,
75014 Paris, France\\ 
$^4$GEPI(CNRS UMR 8111), Observatoire de Paris, 92195 Meudon, France}

\date{Accepted; Received; in original form}

\pagerange{\pageref{firstpage}--\pageref{lastpage}}
\pubyear{2005}
\begin{document}

\maketitle

\label{firstpage}

\begin{abstract}
High-resolution N-body re-simulations of 15 massive ($10^{14}-10^{15}\msol$)
dark matter haloes have been combined with the hybrid galaxy formation model
\gal\ (Hatton et al. 2003), to study the formation and evolution of galaxies
in clusters, within the framework of the hierarchical merging scenario. This
paper describes the high-resolution resimulation technique used to build the
dark matter halo sample, and discusses its reliability.  New features
incorporated in \gal\ include a better description of galaxy positioning
after dark matter halo merger events, a more reliable computation of the
temperature of the inter-galactic medium as a function of redshift, that also
takes into account the reionisation history of the Universe, and a
semi-analytic description of the ram pressure stripping of cold gas from
galactic discs, suffered by galaxies during their motion through the diffuse
hot intra-cluster medium.  Within the multitude of available model results,
we choose to focus here on the luminosity functions, morphological fractions
and colour distributions of galaxies in clusters and in cluster outskirts, at
$z=0$.  No systematic dependency on cluster richness is found either for the
galaxy luminosity functions, morphological mixes, or colour distributions.
Moving from higher density (cluster cores), to lower density environments
(cluster outskirts), we detect a progressive flattening of the luminosity
functions, an increase of the fraction of spirals and a decrease of that of
ellipticals and S0s, and the progressive emergence of a bluer tail in the
distributions of galaxy colours, especially for spirals. As compared to
cluster spirals, early-type galaxies show a flatter luminosity function, and
more homogeneous and redder colours.  An overall good agreement is found
between our results and the observations, particularly in terms of the
cluster luminosity functions and morphological mixes.  However, some
discrepancies are also apparent, with too faint magnitudes of the brightest
cluster members, especially in the B band, and galaxy colours tendentially
too red (or not blue enough) in the model, with respect to the observations.
Finally, ram pressure stripping appears to affect very little our results.
\end{abstract}


\section{Introduction}
\label{sec:intro}
In the most widely accepted model for structure formation, cold dark matter
(CDM) is the dominant component of the Universe. As a consequence, the first
structures that form are small dark matter (DM) haloes, whose subsequent
hierarchical mergers produce larger objects.  Baryons are mixed in smaller
proportions with the DM particles, and their evolution follows and depends on
that of the DM haloes: gas cools and collapses at the centre of the halo
potential wells, thus forming stars and giving rise to the observable
galaxies.  When haloes merge, the galaxies they contain start orbiting in the
new common potential well, thus mutually interacting before they merge with
one another. Therefore, galaxy formation also proceeds in a hierarchical
fashion, and more massive galaxies are the end products of several mergers of
smaller ones.  In such a picture, a galaxy cluster corresponds to a very
massive DM halo, assembled by the hierarchical merger of several smaller
objects, and containing a large number of galaxies that interact with the
diffuse intergalactic medium and among them; field galaxies, instead, are
embedded in much less massive DM haloes, that have suffered fewer merger
events.  Thus, galaxy properties are expected to depend, to a certain extent,
on the mass and the formation history of their host DM haloes.

From an observational point of view, cluster and field galaxies do indeed
show different properties.  While spirals represent the larger fraction of
the galactic population in the field, early-type galaxies prevail in
clusters, and are mainly localised in the central regions (e.g., Dressler
1980; Whitmore, Gilmore \& Jones 1993; Dressler et al. 1997).  Spirals in
clusters are found mainly in the outer regions and they show a significant
deficiency in HI gas with respect to normal disks (e.g., Chamaraux, Balkowski
\& G\'erard 1980; Giovanelli, Chincarini \& Haynes 1981).  Also the fraction
of dwarf galaxies appears to be larger in clusters than in the field, but the
exact dependence of the shape of the galaxy luminosity function on the local
density of the environment is still unclear (for recent discussion, see,
e.g., De Propris et al. 2003; Lin, Mohr \& Stanford 2004, and references
therein).  It is still a matter of debate whether these differences reside in
the conditions existing at the moment when galaxies formed (``nature''
hypothesis), or in their subsequent evolution (``nurture'' hypothesis).
Certainly, dynamical processes like dynamical friction, direct mergers, and
ram pressure stripping do play a role in modifying galaxy properties, and
they should be more efficient in dense environments such as clusters and
groups (see Mamon 1992, 2000).  Major mergers are expected to destroy galaxy
disks and cause spirals to transform into ellipticals. Minor mergers and
rapid encounters, often collectively termed ``harassment'', should heat up
the disk of a galaxy, leading to the creation of a bar that fuels matter from
the disk into the bulge (e.g., Moore, Lake \& Katz 1998).  Moreover, the ram
pressure that the hot intra-cluster gas exerts on the galactic interstellar
medium can lead to the removal of substantial amounts of cold gas from discs
(Gunn \& Gott 1972), thus contributing at modifying their morphology, star
formation rates, luminosities and colours (Biermann \& Shapiro, 1979; Abadi,
Moore, \& Bower 1999; Kenney, van Gorkom, \& Vollmer 2004, and references
therein).

To study galaxy evolution in clusters within the framework of the
hierarchical scenario, and to try and understand how and how much it differs
from that in the field, it is necessary to both follow the merging history of
very massive DM haloes (the hosts of galaxy clusters), and describe the
physical processes that affect galaxy properties in dense environments.  The
approach we have chosen employs the so-called \emph{hybrid} technique
(Roukema et al. 1997; Kauffmann et al. 1999; Diaferio et al. 2001; Springel
et al. 2001b; Mathis et al. 2002; Okamoto \& Nagashima 2001 and 2003; Hatton
et al. 2003; Helly et al. 2003; De Lucia, Kauffmann \& White 2004b), that
combines N-body simulations for following the formation and merging history
of DM haloes, and semi-analytical modelling for describing the physics of the
baryonic component (e.g., Kauffmann, White \& Guiderdoni 1993; Somerville \&
Primack 1999; Cole et al. 2000, and references therein).

Using N-body simulations for the DM component allows one to directly follow
the highly non-linear dynamics of collapses and mergers, and thus to derive
realistic and reliable halo formation and merging histories.  As a drawback,
the mass resolution is usually limited to unsuitably high values: typically,
the particle mass in cosmological N-body simulations is some $10^{10}\msol$,
and only haloes composed of at least 10 particles can be reliably
detected. This means that a hybrid galaxy formation model applied to a
typical N-body simulation allows one to resolve only galaxies more massive
than few $10^{10}\msol$ (assuming a dark-to-light mass ratio of about 10).
Moreover, the smallest galaxies can only be detected, but no information
about their formation history can be obtained, since they derive from the
hierarchical assembly of smaller systems, not resolved in the model.  To
reduce this problem, high-resolution re-simulation techniques exist that
allow an increase, at relatively small computational cost, of the mass
resolution of a given region (for instance, a massive DM halo) selected in a
larger-scale, lower-resolution N-body simulation.

The semi-analytical modelling uses a set of simplified ``recipes'' for
describing the processes that act on the baryonic component within the DM
haloes: gas cooling, star formation, energy feedback from supernovae, galaxy
interactions, etc.  While the physics of these mechanisms is very complex,
the use of simplified prescriptions, formulated on observationally or
phenomenologically motivated basis, has the main advantage of allowing one to
easily modify or switch on/off one given process, and thus to estimate its
role in affecting galaxy properties.  For instance, as we have done here, one
can include in the model the description of ram pressure stripping of cold
gas from galactic discs, and easily study how results change when this
process is taken into account or not.

In the present paper, we study galaxy formation and evolution in clusters by
re-simulating at high resolution a sample of 15 massive DM haloes, and by
using a semi-analytical model to follow the evolution of the baryonic
component.  In Section \ref{sec:simu}, we describe our cluster-size DM haloes
sample and the resimulation technique we have employed to obtain it.  The
package we use to analyse these N-body simulations and to describe the
physical and spectro-photometric evolution of galaxies within the haloes is
called \nobreak{\gal}\ (for \emph{Galaxies In Cosmological Simulations}). It is
described in detail and tested against several observations of the local
Universe in the first paper of this series (Hatton et al. 2003), and it has
been used to reproduce and predict many properties of the Lyman Break
Galaxies at $z\simeq 3$ in Blaizot et al. (2004).  Here, we apply it to the
cluster environment, after introducing slight modifications required by the
improved mass resolution, and taking into account the ram pressure stripping
process. Its main characteristics and the new physical processes we have
introduced in the model are summarised in Section \ref{sec:galics}.  The
results we find in terms of the galaxy luminosity functions, morphological
fractions and colour distributions are presented in Section
\ref{sec:resu}. The analysis mainly aims at studying if and how these
properties vary with cluster richness, and with the local density of the
environment.  Discussions and conclusions are presented in Section
\ref{sec:discuss}. 

In order to facilitate the comparison to model results, observations presented
here have been rescaled using H$_0=70$ km s$^{-1}$ Mpc$^{-1}$.

\section{The cluster-size DM halo sample}
\label{sec:simu}
\subsection{Parent simulation}
\label{sec:VLS}
Under our working hypothesis, a cluster of galaxies is hosted by a very
massive ($10^{14}$--$10^{15}\msol$) DM halo.  With the aim of studying galaxy
formation in clusters with some statistical significance, a sample of 15 DM
haloes in this mass range has been selected from the dissipationless ``Very
Large Simulations'' (VLS; see Yoshida, Sheth, \& Diaferio 2001), where the
comoving volume of the Universe is sufficiently large to contain several
high-mass objects: the box size is $479\,\mpc/h$, with H$_0 = 100\,h^{-1}$ km
s$^{-1}$ Mpc$^{-1}$, and $h=0.7$.  The adopted cosmological model is the
``standard'' $\Lambda$CDM, with $\Omega_m=0.3$, $\Omega_\Lambda=0.7$,
spectral shape $\Gamma = 0.21$, and normalisation to the cluster local
abundance, $\sigma_8=0.9$.  The total number of particles is $512^3$, of
$6.86\, 10^{10}\msol/h$ mass each.

DM haloes have been detected in these simulations by means of the spherical
overdensity criterion (Lacey \& Cole 1994; Tormen, Bouchet \& White 1997),
i.e., they are defined as spheres centred on maximum density peaks in the
particle distribution, with mean density $\rho_\Delta$ as predicted by the
spherical top-hat model: for the adopted $\Lambda$CDM cosmology, $\rho_\Delta
\simeq 97 \,\rho_{\rm crit}$ at $z=0$ (Eke, Cole \& Frenk 1996) , with
$\rho_{\rm crit}$ being the critical density of the Universe at that
redshift.  The radius $\rdelta$ of such spheres thus defines the boundary of
DM haloes, to be distinguished from the \emph{virial} radius $\rvir$, that we
estimate from the measured potential and kinetic energies.  Among all, we
have selected a sample of 15 haloes with total masses $\mdelta$ ranging
between 10$^{14}$ and $2.3\,10^{15}\msol/h$ (see second and third columns of
Table \ref{tab:sample} for the values of $\mdelta$ and $\rdelta$ in our
sample).  They span a variety of shapes, from nearly round to more
elongated. The richness of their environment also changes from case to case,
with the less isolated haloes usually surrounded by pronounced filamentary
structures, containing massive (up to $20\%$ of the selected halo mass)
neighbours.

\begin{table*}
\centering
\caption{Characteristics of the 15 DM haloes in our sample. {\it Columns 2
and 3}: masses (in $\msol/h$) and radii (in Mpc$/h$) of the 15 DM haloes in
the parent VLS simulations, as measured by the spherical overdensity halo
finder ($\Delta\simeq 97$ at $z=0$ for the adopted cosmology); {\it columns 4
and 5}: numbers of high-resolution and low-resolution particles used in the
resimulation runs; {\it columns 6 and 7}: halo masses and radii after
resimulations; {\it columns 8 and 9}: number of galaxies brighter than
$\magb=-17$ and $-19$, respectively, found within the halo virial radius
$\rvir$.}
\begin{tabular}{|l|c|c|l|r|r|l|c|c|l|r|r|}
\hline 
      &         &         &&       &        &&         &         &&        &       \\
Name  & $\mdelta$ & $\rdelta$ &&$\nhr$ & $\nlr$ && $\mdelta$ & $\rdelta$ && $N_{17}$ & $N_{19}$\\
      &         &         &&       &        &&         &         &&        &       \\
\hline 
      &         &         &&       &        &&         &         &&        &       \\
g8    & $2.29~10^{15}$ & 2.72 &&3700120 & 191733 && $2.34~10^{15}$ & 2.75 && 1642& 374\\
g1    & $1.39~10^{15}$ & 2.30 &&2574717 & 202301 && $1.40~10^{15}$ & 2.31 && 1139& 263\\
g696  & $1.30~10^{15}$ & 2.26 &&4870197 & 184314 && $1.14~10^{15}$ & 2.16 &&  901& 235\\
g51   & $1.20~10^{15}$ & 2.20 &&1677364 & 213477 && $1.08~10^{15}$ & 2.12 &&  884& 191\\
g72   & $1.16~10^{15}$ & 2.17 &&3299865 & 194277 && $1.18~10^{15}$ & 2.18 &&  908& 229\\
g245  & $7.08~10^{14}$ & 1.84 &&3437317 & 215269 && $6.50~10^{14}$ & 1.79 &&  487& 103\\
g689  & $5.95~10^{14}$ & 1.74 &&3252085 & 215809 && $6.08~10^{14}$ & 1.75 &&  613& 126\\
g564  & $4.98~10^{14}$ & 1.64 &&2068981 & 227187 && $4.91~10^{14}$ & 1.63 &&  433& 94\\
g1777 & $4.02~10^{14}$ & 1.52 &&3094123 & 219047 && $3.83~10^{14}$ & 1.50 &&  342& 64\\
g4478 & $2.99~10^{14}$ & 1.38 &&2293433 & 225706 && $2.92~10^{14}$ & 1.37 &&  205& 41\\
g6212 & $1.01~10^{14}$ & 0.95 &&271228  & 246780 && $1.12~10^{14}$ & 1.00 &&  92& 23\\
g3344 & $1.01~10^{14}$ & 0.96 &&206140  & 248756 && $1.09~10^{14}$ & 0.99 &&  84& 24\\
g914  & $1.00~10^{14}$ & 0.96 &&250605  & 247091 && $1.45~10^{14}$ & 1.09 &&  52& 10\\
g676  & $1.00~10^{14}$ & 0.96 &&210958  & 248817 && $1.05~10^{14}$ & 0.98 &&  94& 17\\
g1542 & $1.00~10^{14}$ & 0.96 &&207202  & 248948 && $8.22~10^{13}$ & 0.99 &&  78& 15\\
      &         &         &&        &        &&         &         &&        &       \\
\hline
\end{tabular}
\label{tab:sample}
\end{table*}

Given the particle mass of the VLS, only haloes of at least $\sim
7\,10^{11}\msol$ (10 particles) can be detected, and no information about the
formation history of the smallest can be obtained.  This clearly limits the
reliability of any galaxy formation model based on such simulations, since
all galaxies fainter than about $L_*$, which are an important fraction of the
overall galactic population and the building blocks of brighter galaxies,
cannot be resolved.  For that reason, we have re-simulated the 15 haloes of
our sample and increased their mass resolution by a factor $\sim 33$ or more,
as described below (see also Lanzoni, Cappi \& Ciotti 2003; Lanzoni et
al. 2004; De Lucia et al. 2004a and 2004b).

\subsection{High-resolution re-simulations}
\label{sec:HRres}
The technique we use (Tormen et al. 1997) consists in selecting a halo in a
given cosmological simulation, and ``zooming in'' on it.  The first step is
to draw a spherical region around the selected halo in the parent
simulation. The radius of such a region ($\rdec$) is of few $\rdelta$, so
that it contains all the particles composing the halo, plus those in a
boundary layer around it.  The number of particles contained within this
region (that is called the \emph{high resolution region}) in the initial
conditions of the simulation is then increased, until the desired mass
resolution is achieved.  As a consequence, the mean inter-particle separation
decreases, and it becomes necessary to add the corresponding high-frequency
modes of the fluctuation spectrum to the waves on larger scales used in the
initial set up of the parent simulation.  The displacement field
corresponding to these high-frequency modes is also computed and added to
that of the original parent simulation.  The distribution of particles
surrounding the central high resolution (HR) region is smoothed by means of a
spherical grid, whose spacing increases with the distance from the centre: in
such a way, the original surrounding particles are replaced by a smaller
number of \emph{macro-particles}, whose mass grows with the distance from the
HR region.  Thanks to this method, the total number of particles to be
evolved in the simulation remains small enough to require reasonable
computational resources.  At the same time, the tidal field that the overall
particle distribution exerts on the HR region is kept very close to the
original one, and the boundary layer of HR particles around the selected halo
prevents its ``contamination'' by lower-resolution macro-particles during the
simulation. This implies that the boundary layer has to become larger when
the mass of the selected halo decreases, and/or the local density around it
increases, and/or the mass resolution one wants to attain increases.  The
resulting particle distribution in the new initial conditions is spherical,
with a diameter equal to the box size of the parent simulation, and with
vacuum boundary conditions (i.e., a vanishing density fluctuation field
outside the sphere).  An N-body simulation is then run starting from these
new initial conditions, and the properties of the selected halo are
recomputed at the desired higher resolution.

We have applied such a technique to the 15 massive DM haloes selected in the
VLS. To run the resimulations, we have used the parallel dissipationless
tree-code GADGET (Springel, Yoshida \& White 2001a), on the IBM SP2 of the
CINES (Montpellier, France), and the CRAY T3E of the RZG Computing Center
(Munich, Germany).  The resolution in mass has been increased by a factor
$\sim 33$ for the 5 most and the 5 least massive haloes, using
high-resolution particles of $\sim 2\,10^9\msol/h$ each, while a further
factor of 2 in mass resolution was achieved for the 5 intermediate mass
haloes (leading to a particle mass of $10^9\msol/h$ in this case).  As a
consequence, dark matter haloes with masses down to 2 and $1\times
10^{10}\msol/h$ can now be resolved in our resimulations.  The Plummer
equivalent gravitational softening used for the HR region is $\epsilon =
5\,\kpc/h$, corresponding to about $0.2\%$ and $0.5\%$ of the radius
$\rdelta$ for the most and the least massive haloes, respectively. This scale
length, which is a fair estimate of our spatial resolution, is to be compared
with that of $30\,\kpc/h$ in the original VLS.  To prevent low-resolution
macro-particles to end within the final resimulated halo, a radius $\rdec$ of
about 3 to $5\,\rdelta$ was necessary for the most, and the least massive
haloes.  The resulting number of high and low resolution particles used for
the resimulation runs are listed in Table \ref{tab:sample} (columns 4 and 5);
note that to get the same mass resolution in the entire volume of the parent
simulation, about $1650^3$ particles would have been required.  Results have
been dumped in files for 100 temporal outputs, logarithmically spaced in the
expansion factor of the Universe between $z=20$ and $z=0$, yielding a time
resolution of about 400 Myr in the worst case, i.e., around $z=0$.

\begin{figure*}
\centerline{\epsfxsize = 12.6 cm \epsfbox{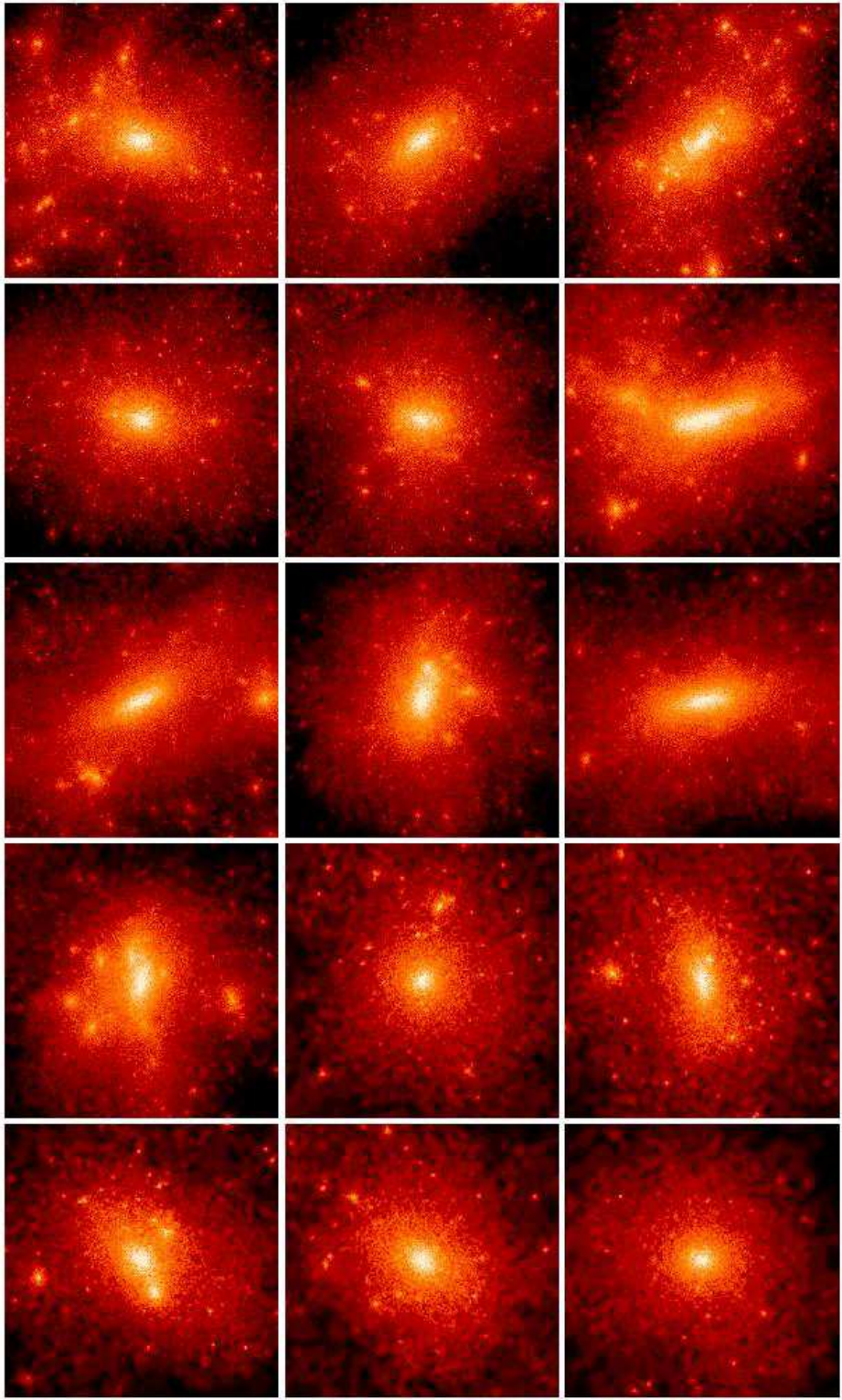}}
\caption{Images of the 15 DM haloes in the high-resolution resimulations at
$z=0$.  Panels show the projection along the $z$-axis of a cubic region of
about $2\rdelta$ side centred on each halo. Particle density increases from
black to white. Haloes are ordered from left to right, then top to bottom as
listed in Table \ref{tab:sample} (decreasing order of mass in the
low-resolution simulations).}
\label{fig:ima}
\end{figure*}

Images of the 15 haloes after resimulations are shown in Figure
\ref{fig:ima}.  Their masses and radii are listed in columns 6 and 7 of Table
\ref{tab:sample}. They differ by a few percent from the original values
(columns 2 and 3 of the same table), because the overall gravitational field
in the resimulations is not exactly identical to the original one in the VLS.
These differences are smaller than $10\%$ in mass and $4\%$ in radius for the
10 most massive haloes, while they are slightly larger for the lighter
objects.  Halo g914 represents an extreme case: its high-resolution mass and
radius are respectively $45\%$ and $13\%$ larger than its low resolution
values.  This can be explained by noticing that, in the parent simulation,
g914 had a neighbouring halo with a mass of $\sim 2\,10^{13}\msol/h$ (equal
to $20\%$ of its mass), at a distance of about $2.5\,\rdelta$.  This object
is not found anymore in the resimulation, where, instead, two haloes of $1.5$
and $4.5\,10^{12}\msol/h$ are found at $3\,\rdelta$, and another one of
$1.5\,10^{13}\msol/h$ is at a distance of $3.5\,\rdelta$.  It is therefore
possible that, due to the lack of a massive neighbour in the resimulation,
g914 has been able to accrete more matter, resulting in a mass and radius
increase by an important fraction.

In summary, we can assess that the high-resolution re-simulation technique is
reliable, in the sense that it reproduces quite accurately the selected halo
properties (and those of a number of other objects within the boundary layer
around them) at increased mass resolution (see also Gao et al. 2004).  Thanks
to our 15 resimulations, we are therefore able to describe galaxy formation
and evolution in clusters and in their outskirts in much more detail than we
would if we were using the VLS alone.

\section{The GalICS galaxy formation model}
\label{sec:galics}

\subsection{General overview}
\label{sec:overview}
Once the N-body simulations for the DM component have been run, we have used
the \gal\ hybrid galaxy formation model to study galaxy formation and
evolution within the haloes.  The model consists in three main steps: 1)
detecting DM haloes in the outputs of the N-body simulation; 2) constructing
their merging history trees; 3) describing the physics of the baryonic
component within the haloes by means of a semi-analytical technique (e.g.,
White \& Frenk 1991; Kauffmann et al. 1993; Cole et al. 1994; Somerville \&
Primack 1999). Its main features are described below, but we refer to Hatton
et al. (2003) for full details.

To identify DM haloes in the outputs of our high resolution resimulations, we
use a \emph{friends-of-friends} algorithm (Davis et al. 1985), with
linking-length parameter $b=0.2$, in units of the mean inter-particle
separation.  Only haloes with at least 10 particles (i.e., $\sim 1-2\times
10^{10}\msol/h$) are detected, since smaller groups are usually dynamically
unstable (Kauffmann et al. 1999).

Haloes of each timestep are then linked to those in the preceding and in the
successive outputs, by following the DM particles they have in common.  A
sort of genealogical tree is therefore built, where parents and descendants
are recorded. Events of diffuse matter accretion and halo fragmentation can
also take place and contribute to increase/decrease the mass of the resulting
halo.

Galaxy properties are then computed and evolved through the DM halo merging
tree, by means of simplified ``recipes'', physically or phenomenologically
motivated, that are able to describe the processes acting on gas and stars.
The overall scheme of our semi-analytical model is similar to that of others:
\begin{itemize}
\item gas cools and forms stars in an exponential, rotationally supported
disc at the centre of newly detected DM haloes, provided that they are bound
and that their spin parameter $\lambda < 0.5$;
\item supernovae explosions reheat the gas and possibly eject it out from the
disc or the halo, thus inhibiting future star formation, and spreading metals
into the interstellar medium; 
\item when haloes merge, the galaxies they contain start orbiting in one
common potential well, and dynamical interactions take place: orbital decay,
due to dynamical friction against the background DM particles, makes galaxies
sink towards the centre of the halo where they eventually merge with the
central galaxy; direct mergers can also take place between satellite (non
central) galaxies.
\end{itemize}
Within this scheme, morphological transformations from discs to spheroids are
driven by merger events, with an efficiency that depends on the initial mass
ratio of the interacting galaxies. Note that disc instabilities are also
taken into account as possible mechanisms for feeding central spheroidal
components.  New discs of cold star-forming gas can be later accreted by the
spheroids, and spiral galaxies, with a more or less prominent bulge, thus
form.  Stellar spectra are computed along the full merging history of
galaxies by means of a spectrophotometric model, so that galaxy luminosities
and colours are obtained at every timestep.  For this purpose, we use
STARDUST (Devriendt, Guiderdoni \& Sadat 1999), which also takes into account
the effects of dust absorption and emission.

As in \galone, galaxies are morphologically classified by their bulge-to-disc
luminosity ratio in the B band. Following Simien \& de Vaucouleurs (1986), we
define a morphological index $I\equiv\exp(-L_B/L_D)$, so that $I=1$ for a
pure disc, $I=0$ for a pure bulge, and the different morphological types
correspond to the following intervals of it: $I\le 0.219$ for elliptical
galaxies (corresponding to a bulge blue light fraction, with respect to the
total, $> 60\%$); $0.219 < I < 0.507$ for lenticulars (bulge fraction between
40\% and 60\%); $I \ge 0.507$ for spirals (bulge fraction smaller than 40\%).
The values adopted for the free parameters are also the same as those used in
\galone\ (see their Table 2).

\subsection{New elements in the GalICS model}
\label{sec:new_ele}
With respect to the original version of the model as presented in \galone,
several modifications have been introduced in the present work.

The first change concerns galaxy positioning: by using equation (5.1) of
\galone\ to place galaxies within the new common potential well after DM halo
merger events, we noticed that many of our most massive clusters did not have
any central galaxy, at odds with observations. We have therefore slightly
modified that prescription by weighting the original `jumping' distance $R_j$
of \galone\ by the mass contribution of the progenitor halo to its son:
$R_{j, {\rm new}} = R_j\,(1-M_{\rm prog }/M_{\rm son})$. In this way, if the
progenitor contributes to most of the mass of the son, its position (and that
of its central galaxy) remains practically unperturbed by the merger event,
while if it is negligible, the new position of its central galaxy is
determined as in \galone.  It has been shown (Springel et al. 2001b) that
correctly following the spatial distribution of galaxies within DM haloes,
and thus also their merging rate, is quite important for getting the right
galaxy properties, like the bright-end of the luminosity function, or the
morphology-clustercentric relation.  Whereas a substantial improvement in
this direction will be implemented in the next version of \gal, the method we
adopt here is still simplistic and approximate, and we therefore do not
attempt to reproduce in detail those observations that strongly depend on
galaxy radial positions within the clusters.

Second, given the improved mass resolution of our resimulations with respect
to \galone\ (thus, also a large number of low mass haloes that were not
resolved there), a more accurate description of the reionisation history of
the Universe is necessary to avoid an unobserved excess of galaxies at the
faint-end of the luminosity function (see thin solid lines in Figure
\ref{fig:4lf}). For that purpose, here we compute the temperature of the
inter-galactic medium as a function of redshift following Valageas \& Silk
(1999), and we use it as a threshold to determine whether the gas can cool
and form galaxies within DM haloes. More specifically, we require that the
temperature of the infalling gas (taking into account adiabatic heating as it
falls in the DM potential well) be smaller than the virial temperature of the
DM halo if the gas is to cool and form a central galaxy (e.g., Blanchard,
Valls-Gabaud \& Mamon 1992).

Finally, we include in this work the description of ram pressure stripping of
cold gas from galactic discs, as they orbit through the hot intra-cluster
medium.  This process is thought to be responsible for the neutral hydrogen
deficiency observed in cluster spirals with respect to those in the field
(Gunn \& Gott 1972), and has also been considered as a possible mechanism to
turn disc galaxies into lenticulars (e.g., Melnick \& Sargent 1977; Solanes
\& Salvador-Sol\'e 1992).  Despite its potential importance, such a process
has only been implemented once in a hybrid model for galaxy formation
(Okamoto \& Nagashima 2003). This previous implementation led to the
conclusion that ram pressure stripping produces negligible changes in galaxy
colours, star formation rates and morphological fractions within the cluster
core. These results, however, were obtained only for bright galaxies
($L>L_*$) in one single cluster, while the present work allows us to extend
the investigation to fainter galaxies and a larger number of clusters of
different masses.  

We have modelled gas stripping from discs following the original idea of Gunn
\& Gott (1972), who use a simple argument of equilibrium between static
forces to estimate its efficiency: if the dynamical pressure due to the
intra-cluster medium is larger than the gravitational force per unit surface
of the disc, all cold gas beyond a given radius $R_{\rm str}$ on the disc is
stripped out.  For an exponential density profile, such a radius is given by:
\begin{equation}
\frac{R_{\rm str}}{R_{\rm D}} = -\ln\sqrt{\frac{\rho_{\rm ICM}(r)\,v^2_\perp}
{2\,\pi\,G\,\Sigma_{0\star{\rm g}}\,\Sigma_{0\rm g}}},
\label{eq:rstrip}
\end{equation}
where $R_{\rm D}$ is the characteristic radius of the gas and of the
gas-plus-stars exponential distributions in the disc, with central surface
brightness $\Sigma_{0\rm g}$ and $\Sigma_{0\star{\rm g}}$, respectively;
$\rho_{\rm ICM}(r)$ is the intra-cluster medium density at the orbital
position $r$ of the galaxy within the cluster at the beginning of each
time-step; $v_\perp$ is the component of the galaxy velocity perpendicular to
the disc.  Further assuming that the distribution of galaxy orbital
velocities is a Maxwellian, with a rms velocity equal to the halo velocity
dispersion (i.e., $\sigma=\vcirc/\sqrt{2}$, for a singular isothermal sphere
of circular velocity $\vcirc$), one can randomly select the module of
$v_\perp$.  One then multiplies it by the sine of the angle between the
velocity vector and the disc plane, this time randomly selected from a
uniform distribution between 0 and 1, to finally determine $R_{\rm str}$.
The mass of cold gas beyond this stripping radius is then removed from the
disc and added to the hot intra-cluster medium.

\subsection{Practical considerations}
\label{sec:prac}
We have applied the (slightly modified) \gal\ model to the 15 high-resolution
resimulations of massive haloes presented above.  Since the halo masses span
one order of magnitude, from $10^{14}$ to $10^{15}\msol/h$, we have studied
the dependence of galaxy properties on cluster richness (mass).  Moreover,
not only the main clusters, but also their outskirts were available at high
resolution in every simulation; therefore, we also investigated the
dependence of galaxy properties on the local environment. In particular, we
distinguish among the ``{\it whole cluster}'', i.e., the entire DM halo and
its baryonic content within a sphere of radius $\rvir$; the ``{\it cluster
core}'', i.e., DM and baryons contained within a sphere of radius $\rvir/3$;
the ``{\it cluster envelope}'', i.e., the region between $\rvir/2$ and
$\rvir$; and the ``{\it cluster outskirts}'', i.e., all high-resolution DM
haloes less massive than $10^{14}\,\msol/h$ around the main cluster (within
the high-resolution boundary layer), and their baryonic content.  Note
however that, our definitions of cluster ``core'' and ``envelope'' do not
exactly match the observed ones, since they refer to 3-dimensional volumes,
while observations always deal with projections along the line of sight.

Random realisations are used in some part of the semi-analytical code to
determine various galaxy properties.  Besides those used for the ram pressure
stripping (see previous section), we use random numbers also for assigning
galaxy positions within the new common potential well after DM halo mergers,
and for selecting the satellite galaxies involved in direct collisions.  To
check how much results are sensitive to random realisations, we have run the
model 10 times on the same simulation, only changing the value of random
seeds.  We find that results are in general very robust, with a small
dispersion about the mean values.

As discussed in \galone, the existence of a cut-off mass for the DM haloes
(10 particles, in our case) introduces an effective resolution limit on our
results.  We therefore take as ``completeness limit'' for the baryonic mass
of our galaxy sample the smallest halo mass times the baryonic fraction of
the Universe, i.e., the galactic mass
$m_{lim}=2\,10^{10}\times\Omega_b/\Omega_m\simeq 4\, 10^9\,\msol$, where the
value $\Omega_b=0.02/h^2$ is assumed (Tytler, Fan \& Burles 1996). The tight
correlation that is found between galaxy baryonic mass and magnitudes, gives
in turn the following magnitude limits in the B, V, and K bands: $\magb\simeq
-16.75$, $\magv\simeq -17.5$, and $\magk\simeq -20.75$. Galaxies fainter than
these limiting magnitudes are never considered in the following.

\section{Results}
\label{sec:resu}

\subsection{Luminosity functions}
\label{sec:LF}
\begin{figure}
\centerline{\epsfxsize = 8.7 cm \epsfbox{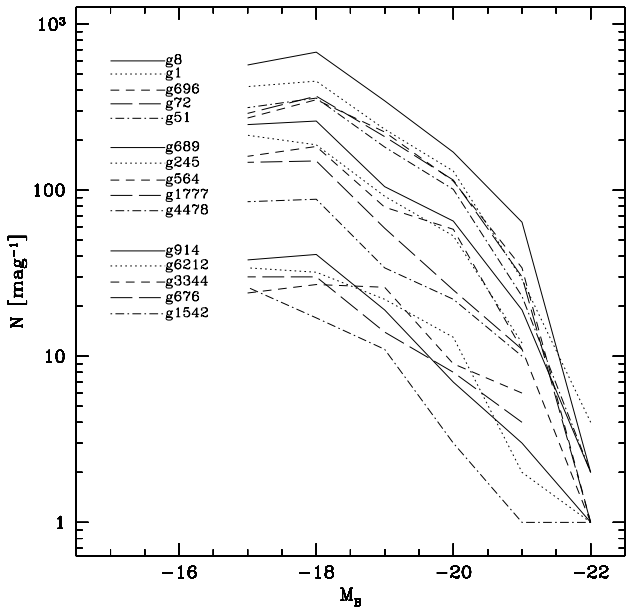}}
\caption{Dependence of the galaxy luminosity function on cluster mass.  The B
band LFs of our 15 simulated clusters are plotted in units of number of
galaxies per magnitude bin. Curves are labelled with halo names, the halo
mass approximately increasing along the y-axis (compare to Table
\ref{tab:sample}).}
\label{fig:single_lf}
\end{figure}

The individual B band galaxy luminosity functions (LFs) of the 15 simulated
clusters are plotted in Figure \ref{fig:single_lf}.  The shape of the LF
appears to vary from cluster to cluster (which is reminiscent of, e.g., the
observations of Yagi et al. 2002), but no systematic trend is detected as the
halo masses change.

\begin{figure*}
\centerline{\epsfxsize = 17.6 cm \epsfbox{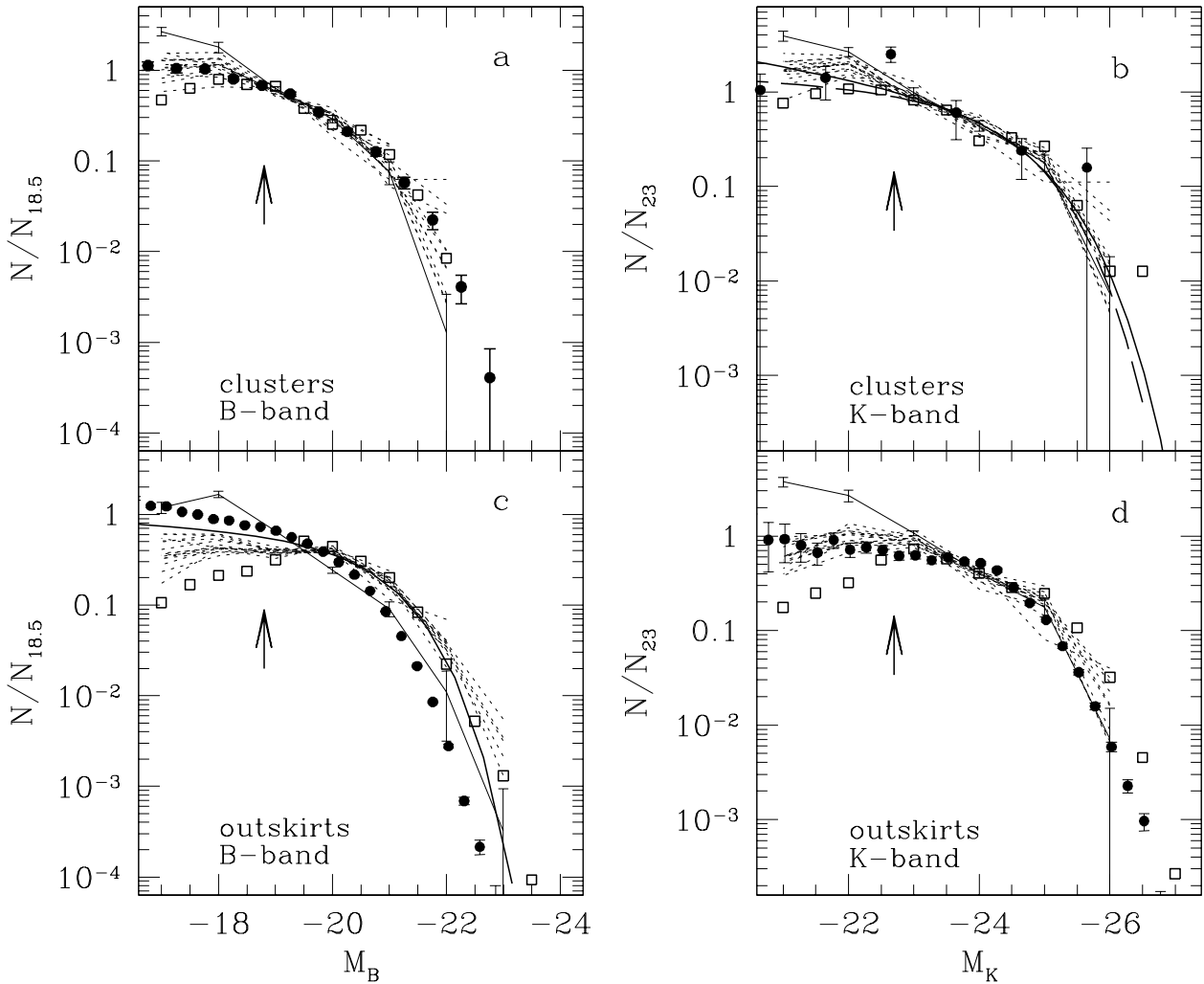}}
\caption{Comparison of the model LF to observations, for galaxies in clusters
and in cluster outskirts, in the B and K bands. All LFs are normalised to the
number of galaxies brighter than $\magb=-18.5$ or $\magk=-23$.  In all
panels, dotted lines show the model LFs for each individual cluster.  Thin
solid lines mark the average LF of the 15 clusters when the reionisation
history of the Universe is not taken into account. Empty squares trace the LF
determined from \galone\ (error bars are omitted for sake of clarity), for
the galaxies populating the most massive DM halo ({\it panels a} and {\it
b}), and the average field ({\it panels c} and {\it d}); the magnitude
resolution limits of \galone\ are marked with the arrows.  {\it Panel a:} B
band LFs of cluster galaxies.  Solid circles represent the composite B-band
LF of 60 clusters from the 2dFGRS (De Propris et al. 2003; the conversion
from $b_j$ to Johnson B is done as described in Table~2).  {\it Panel b:} K
band LFs of cluster galaxies.  Solid circles represent the composite LF of 5
X-ray luminous clusters at various redshifts, as determined by Trentham \&
Mobasher (1998). The dashed and the solid lines are, respectively, the
Schechter best-fits to the composite K band LF of clusters and groups from
the 2MASS Survey by Lin et al. (2004), and from the 2MASS and LCRS surveys,
as determined by Balogh et al. (2001).  {\it Panel c:} B band LF of galaxies
in cluster outskirts and in the field.  Solid circles represent the field LF
from 2dFGRS (Norberg et al. 2002), after the conversion from $b_j$ to Johnson
B magnitudes. The solid line trace the B band LF of Marinoni et al. (1999).
{\it Panel d:} K band LF of galaxies in cluster outskirts and in the field.
Solid circles represent the field LF from the 2dFGRS (Cole et al. 2001).  See
Table~2 for the values of the best-fit Schechter parameters.}
\label{fig:4lf}
\end{figure*}

The self-similarity of the cluster LF is even clearer when normalising to the
number of galaxies brighter than a given magnitude, as in Figure
\ref{fig:4lf} (upper panels), for $\magb\le-18.5$ and for $\magk\le-23$: the
15 dotted lines are quite closely superimposed, and minor deviations from the
average LF are independent of cluster mass.  Also the LFs of the most massive
cluster in \galone\ ($3.9\,10^{14}\msol/h$) is well superimposed on those of
the present 15 clusters, both in B and in K, above the completeness limits of
that work (marked with arrows in the figure).  We therefore conclude that,
\emph{at least in the considered cluster mass range ($10^{14}\div
10^{15}\msol$), and for $\magb\lsim -17$ and $\magk\lsim -21$, the galaxy
luminosity function in clusters is universal}.  This is also the conclusion
of several observational studies in different photometric bands (e.g.,
Trentham 1998; Trentham \& Mobasher 1998; Paolillo et al. 2001; De Propris et
al. 2003), even if a general consensus still lacks (see, e.g., Lopez-Cruz et
al. 1997; Lumsden et al. 1992; Valotto et al. 1997; Trentham \& Hodgkin 2002;
Lin et al. 2004). In particular, Trentham \& Hodgkin (2002) detect a
steepening of the faint-end of the LF for increasing cluster richness, but
this trend is clear only at magnitudes fainter than $\magb=-17$, where our
simulations start to become incomplete.  In the K band, a (mild) steepening
of the faint-end LF for decreasing cluster richness, is found both by Lin et
al. (2004) and in our model, but the statistical significance is poor.  A
more robust difference between rich and poor clusters seems to be the
observed fainter luminosity of the brightest galaxies in lower mass clusters,
compared to the most massive ones (De Propris et al. 2003; Lin et al. 2004),
a trend that we also find in our model.

Given its apparent universality, we can compare our results to the observed
\emph{composite} LFs, obtained by combining together clusters of different
richness.  Figure \ref{fig:4lf} (panel a) shows that a remarkable agreement
is found between the model B band LF and that obtained from 60 clusters in
the 2dF Galaxy Redshift Survey (De Propris et al. 2003), which very well
follows, in turn, the one derived by Trentham (1998) for 9 clusters of
different richness at $z<0.2$.  Also in K (panel b), model results agree well
with the composite LF of Trentham \& Mobasher (1998) derived for 5 X-ray
luminous clusters at different redshifts, that of Balogh et al. (2001) from
the 2MASS and LCRS surveys, and that of Lin et al. (2004) obtained for
clusters and groups from the 2MASS survey.  Actually, in both bands a
discrepancy is also apparent: galaxies brighter than $\magb\simeq -22$ and
$\magk\simeq -26$ are absent in the model, but are found in the observations.
Note, however, that these very bright objects are also the rarest, and we
might miss them because of insufficient statistics.

The model LFs in cluster outskirts are shown in the lower panels of Figure
\ref{fig:4lf}. The LFs of \galone\ for the average field, both in B and in K
bands, are in rather good agreement with those of our cluster outskirts,
except at the faint-end, where the effects of the lower mass resolution of
\galone\ rapidly become apparent. With respect to the observations (Marinoni
et al. 1999 and Norberg et al. 2002, for the B band; Cole et al. 2001, in K),
the overall agreement is less satisfactory than for clusters.  Also in this
case, the brightest observed galaxies are missed in our simulations. However,
they are found in \galone, where the model is run on a scale supposedly
representative of the ``average field'' of the Universe.  This suggests that
such a discrepancy is not a drawback of the model itself, but it is due to
the fact that we are not comparing exactly the same kinds of environment
here, since only the immediate outskirts of clusters, not the average field,
are available in our resimulations.  This same reason might also account, at
least in part, for the too bright characteristic magnitude $\mstar$ of our
LFs, with respect to the observations: since $\mstar$ is typically fainter in
clusters, than in the field (see below), and since observational samples of
the field also include data from galaxies residing in clusters, a fainter
$\mstar$ for the field, than for the immediate cluster outskirts, has to be
expected. In addition, as apparent from panel c, the entity of such a
disagreement also depends on the data sample we compare with, since
inconsistencies are still present among different observational works (for
example, those between Marinoni et al. 1999 and Norberg et al. 2002, which
cannot be solely attributable to uncertain transformations between their
photometric systems. See, e.g., Liske et al. 2003 for an extensive discussion
of disagreements among different LFs).

For a more quantitative comparison, we have fitted our LFs with the Schechter
(1976) function. For magnitudes above the completeness limits, binned by
half-magnitude, the Schechter parameters obtained by minimising the $\chi^2$
of the joint fits to the 15 cluster LFs are given in Table~2, where we also
list those derived by some observational works. Within the uncertainties, the
Schechter parameters of the model and the observed cluster LFs agree well.
Instead, with respect to the B band observations in the field, $\mstar$ is
0.13--0.78 mag too bright (depending on the observational sample of
comparison), and the slope is too flat in our simulations. The ``correct''
slope is found with respect to the observed field LF in K, but an excess of
0.2 mag in $\mstar$ is also apparent.  In any case, it is worth noticing that
the values of the best fit parameters are quite dependent on several factors,
like whether the brightest cluster members are included (as in our case) or
not in the analysis (e.g., Lin et al. 2004), and the considered range of
magnitudes (for instance, if we restrict the fits to galaxies 0.5 mag
brighter than the limiting magnitudes, we find steeper faint-end slopes).
Moreover, a single Schechter function often provides a poor fit to the
observed LFs (e.g., Popesso, Boehringer \& Voges 2004, and references
therein), and no firm agreement is yet found among different observational
studies.
\begin{table}
\begin{center}
\caption {Values of the best-fit Schechter parameters and their $1\,\sigma$
error bars for the model and the observed LFs, in B and K bands. The upper
part of the table refers to cluster environment, the lower part to the
cluster outskirts for the model, and to the ``average field'' of the Universe
for the observations.}  
\renewcommand{\arraystretch}{1.75} \tabcolsep 2.2pt
\begin{tabular}{lcccll}
\hline 
Band  & envmt & input & ref. & \multicolumn{1}{c}{$\mstar$} & \multicolumn{1}{c}{$\alpha$} \\
\hline 
B & cluster & simulations & {\sc g6} & $-20.56^{+0.04}_{-0.10}$ & $-1.34^{+0.01}_{-0.03}$ \\
K & cluster & simulations & {\sc g6} & $-24.75^{+0.07}_{-0.04}$ & $-1.39^{+0.01}_{-0.01}$ \\
B & cluster & {\scriptsize APM+2dFGRS} & DP03 & $-20.58\pm0.07$ & $-1.28\pm0.03$ \\
K & cluster & {\scriptsize 2MASS+LCRS} & B01 & $-24.58\pm0.40$  & $-1.30\pm0.43$ \\
K & cluster & {\scriptsize 2MASS+2dFGRS} & L04 & $-24.34\pm0.01$ & $-1.1$ \\
\hline
B & outskirts & simulations & {\sc g6} & $-20.95^{+0.01}_{-0.05}$ & $-0.98^{+0.01}_{-0.01}$ \\
K & outskirts & simulations  & {\sc g6} & $-24.41^{+0.02}_{-0.07}$ &$-0.98^{+0.02}_{-0.02}$ \\
B & field & {\scriptsize NOG} & M99 & $-20.82\pm0.08$ & $-1.1\pm0.06$ \\
B & field & {\scriptsize APM+2dFGRS} & N02 & $-20.17\pm0.07$ & $-1.21\pm0.03$ \\
K & field & {\scriptsize 2MASS+2dFGRS} & Co01 & $-24.21\pm0.03$ & $-0.96\pm0.05$ \\
\hline
\end{tabular}
\end{center}
Notes: envmt = environment; refs. = references: {\sc g6} = 
{\sc galics vi} (this paper); 
DP03 = De Propris et al. (2003); 
B01 = Balogh et al. (2001);
L04 = Lin et al. (2004), obtained by fixing $\alpha=-1.1$;
M99 = Marinoni et al. (1999); 
N02 = Norberg et al. (2002); 
Co01 = Cole et al. (2001).  
The value of $\mstar$ of DP03 and N02 is corrected for the conversion from
$b_J$ to Johnson B, by making use of $b_J - $B $= -0.28$ (B--V) (Blair \&
Gilmore 1982) and $\langle$B--V$\rangle \simeq 0.94$ (Norberg et al. 2002).
\label{tab:fit}
\end{table}

\begin{figure}
\centerline{\epsfxsize = 8.7 cm \epsfbox{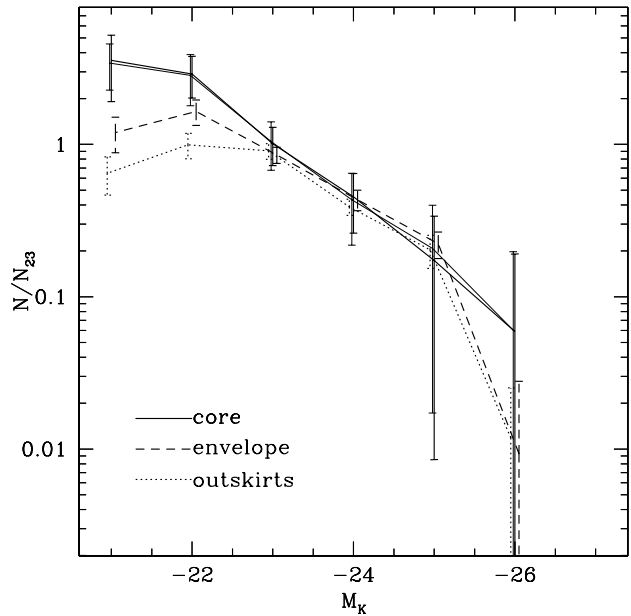}}
\caption{Dependence of the K band LF on environment. Curves and error bars
represent the average LFs of the 15 simulations, and their rms dispersion
about the mean, for three types of environment: cluster cores (thick solid
line), cluster envelopes (dashed line), cluster outskirts (dotted line). The
thin solid line traces the average LF for cluster cores when ram pressure
stripping is neglected. Analogous trends are found in the B band. LFs are
normalised to the number of galaxies with $\magk\le-23$.}
\label{fig:lf_env}
\end{figure}

The dependence of the LF on environment is further studied in Figure
\ref{fig:lf_env}, which reveals a \emph{gradual flattening of the faint-end,
when moving from denser (cluster cores), to less dense regions (cluster
outskirts), a trend that we find both in K and in B bands}. 
Observationally, a steeper faint-end slope of the LF in clusters, and
particularly in cluster cores, with respect to the field, is reported by
several authors, in different photometric bands (e.g., Balogh et al. 2001;
Yagi et al. 2002; De Propris et al. 2003), even if a general consensus still
lacks (Lobo et al. 1997; Trentham 1998; Phillipps et al. 1998; Beijersbergen
et al. 2002; Mobasher et al. 2003; Chrislein \& Zabludoff 2003; Lin et
al. 2004) and conclusions also depend on the considered range of magnitudes.

The model predicts that while the most luminous ``red'' galaxies reside in
clusters cores, the brightest ``blue'' galaxies ($\magb<-22$) populate the
field, but are not seen in clusters (compare panel a and c of Figure
\ref{fig:4lf}).  This latter finding is in agreement with the observations of
Trentham (1998), but in contrast with De Propris et al. (2003); in both
cases, however, observational evidences are not very strong.  De Propris et
al. also report on an excess of very bright ``blue'' galaxies in cluster
cores with respect to the envelopes, which we do not detect in our
simulations. This might be an effect of the too rough way we assign positions
to galaxies within DM haloes (see Section \ref{sec:new_ele}), and not a real
problem of how galaxy formation is modelled, and we plan to investigate this
issue in more detail later.

\begin{figure}
\centerline{\epsfxsize = 8.7 cm \epsfbox{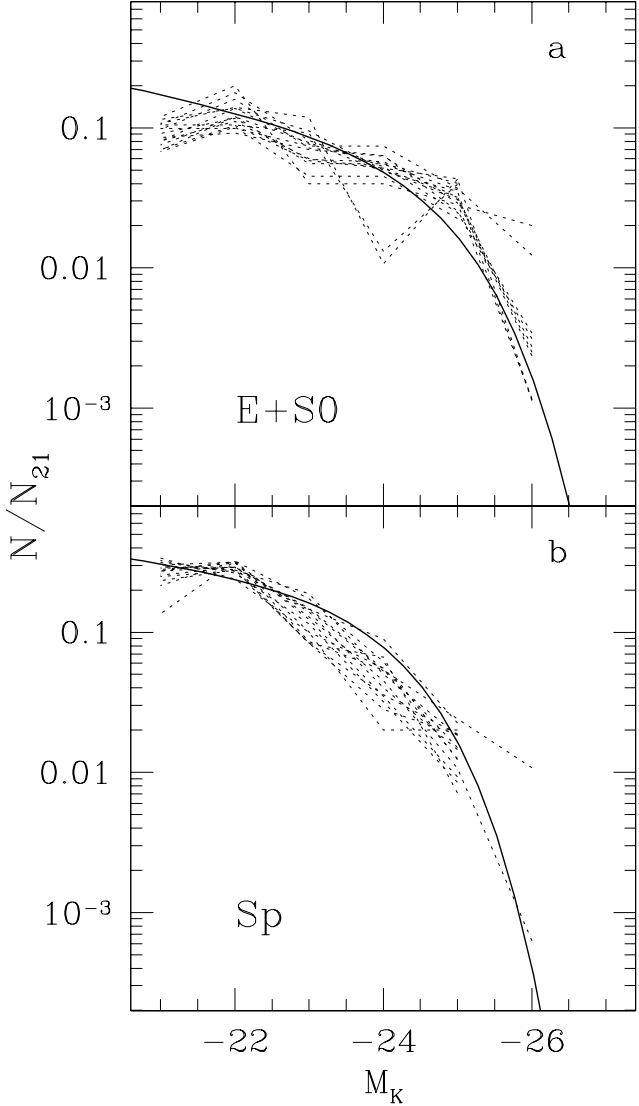}}
\caption{Dependence of the cluster K band LF on morphological types.  Dotted
lines show the 15 individual LFs for model ellipticals and lenticulars ({\it
panel a}), and for model spirals ({\it panel b}), normalised to the total
number of cluster galaxies with $\magk\le-21$. Solid lines are the observed
best-fit Schechter functions, with the same normalisation, as determined by
Balogh et al. (2001) for non-emission-line and emission-line galaxies,
respectively.}
\label{fig:lf_mt}
\end{figure}

In agreement with the observations (Balogh et al. 2001; Yagi et al. 2002; De
Propris et al. 2003), we find that \emph{the LF of early-type galaxies has a
flatter faint-end slope and a brighter characteristic magnitude, with respect
to that of spirals, both in clusters and in the field}. This can be seen in
Figure \ref{fig:lf_mt} for cluster environment and the K band, with model
results for early and late type galaxies compared to the best-fit Schechter
functions for non-emission-line and emission-line galaxies, as determined by
Balogh et al. (2001) on the basis of the measured equivalent width of the [O
II] emission line.  We emphasize that the normalisation is {\em not}
arbitrary in Figure \ref{fig:lf_mt}: both model and observed LFs are
normalised to the {\em total} number of cluster galaxies with $\magk\le-21$.
A similar agreement is found with the data of De Propris et al. (2003) in the
b$_j$ band, for galaxies types split following their spectral properties
(Madgwick et al. 2002).  This should be interpreted as a non trivial success
of the model, particularly if one considers that our definition of
morphological types (based on the bulge-to-disc luminosity in the B band)
does not exactly correspond to those adopted in the observational studies.

\subsection{Morphological fractions}
\label{sec:mt}
\begin{figure}
\centerline{\epsfxsize = 8.7 cm \epsfbox{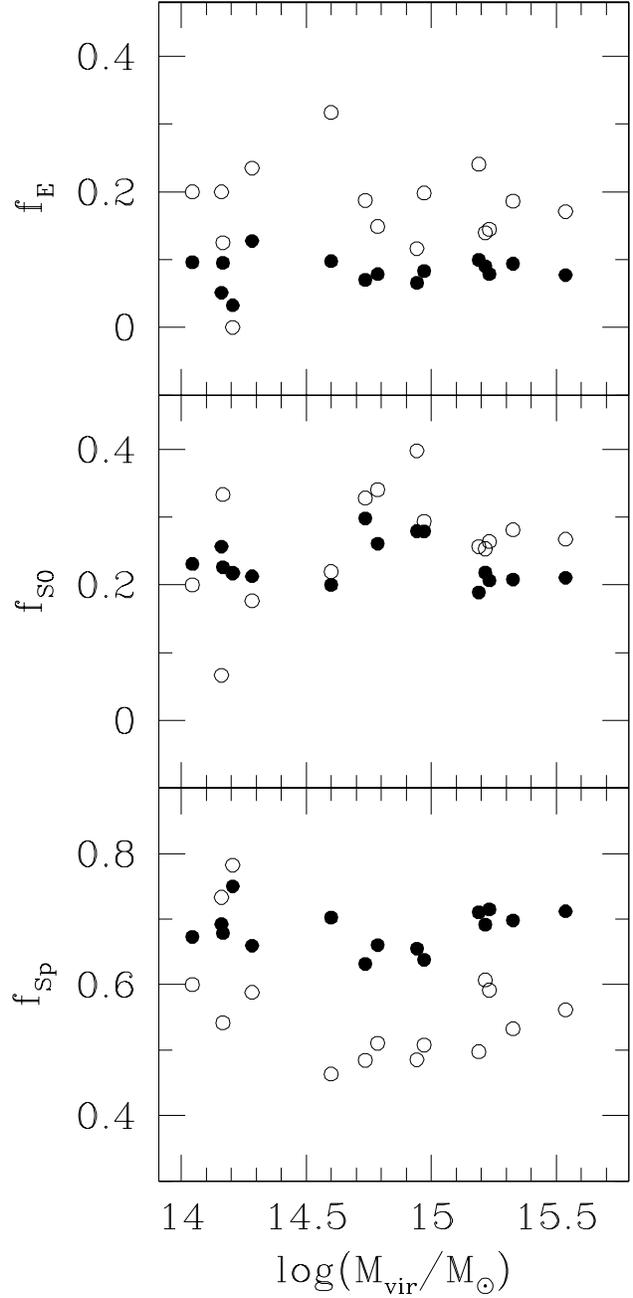}}
\caption{Dependence of galaxy morphological fractions on cluster mass.  From
top to bottom, the 3 panels show the fraction of ellipticals, lenticulars and
spirals, as a function of the host DM halo virial mass. All galaxies within
$\rvir$, having $\magb<-17$ (solid circles) and $\magb<-19$ (empty circles)
are considered.}
\label{fig:morph_richn}
\end{figure}

The dependence of the mix of morphological types on cluster mass for our
simulations is studied in Figure \ref{fig:morph_richn}, for galaxies brighter
than $\magb=-17$ and $\magb=-19$, and for morphologies modelled by the
bulge-to-disk blue luminosity ratio (Sect.~\ref{sec:overview}). The total
number of galaxies brighter than these magnitude limits in each cluster is
given in Table \ref{tab:sample} (columns 8 and 9). From Figure
\ref{fig:morph_richn} we conclude that \emph{the fractions of different
morphological types do not show any clear systematic dependence on cluster
richness.}  The only trend is for a larger fraction, on average, of bright
($\magb<-19$) spiral galaxies in low-mass clusters, but the significance is
marginal.

\begin{figure}
\centerline{\epsfxsize = 8.7 cm \epsfbox{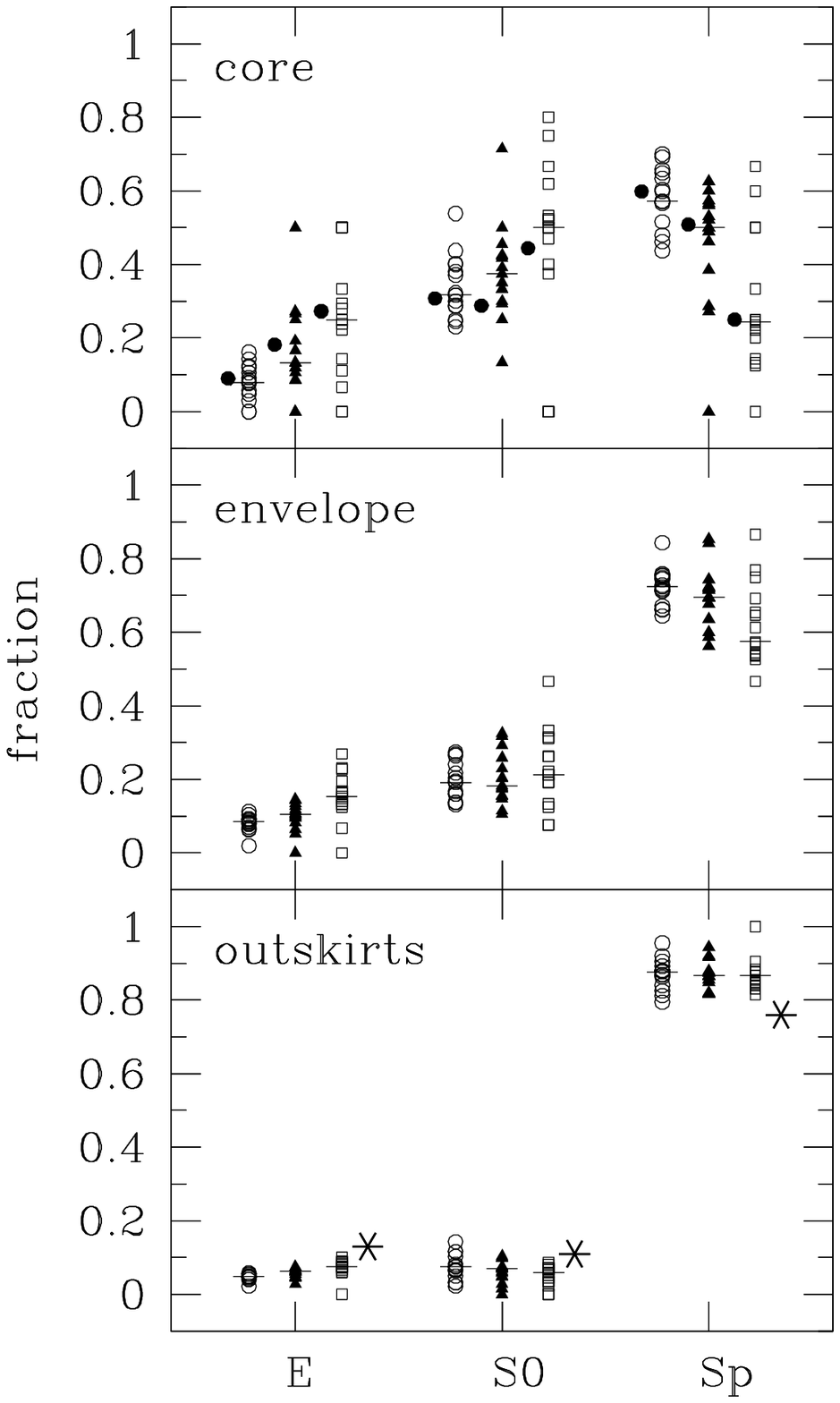}}
\caption{Dependence of galaxy morphological fractions on environment.  The
fractions of ellipticals, S0s and spirals are shown for three kinds of
environment and for three magnitude selection limits. {\it From top to
bottom:} cluster cores, envelopes, and outskirts.  Empty circles are for
galaxies brighter than $\magb=-17$, solid triangles for $\magb<-18$, and
empty squares for $\magb<-19$. The short horizontal lines mark the medians of
the distributions. Solid dots in the upper panel show the median values
obtained if ram pressure stripping is not taken into account.  Asterisks in
the bottom panel mark the average morphological fractions in the field
derived from the Stromlo-APM redshift survey, for galaxies brighter than
$\magb=-18.8$ (E:S0:Sp = 13:11:76).}
\label{fig:morph_env}
\end{figure}

On the other hand, the dependence on the luminosity cut is stronger: with
respect to the global population, the brighter sub-sample shows a higher
fraction of ellipticals and a lower fraction of spirals, independent of
cluster mass.  This is particularly significant in high-density environments,
as apparent from Figure \ref{fig:morph_env}, where the cores, envelopes and
outskirts of the 15 simulated clusters are considered separately: outside of
clusters, the percentage of different morphological types is fairly
independent of the considered cut in absolute magnitude, whereas changes in
the luminosity selection limit lead to increasingly different morphological
mixes when one moves towards the cluster cores.  In other words, \emph{the
relative morphological mix in the inner regions shows larger variations from
cluster to cluster and strongly depends on the magnitude limit, with spiral
galaxies dominating for $\magb<-17$, and ellipticals and S0s catching up at
$\magb<-19$}.  In comparison, \emph{in the outskirts of clusters, the
prevalence of spirals is overwhelming for any luminosity cut with the
percentage of early-type galaxies never in excess of 20\%.}  In the
intermediate density environment of cluster envelopes, late-type galaxies
tend to dominate the overall population, even if their median fraction
systematically decreases for increasing luminosity cut.
%

Our spiral fraction in cluster outskirts appears too high, in comparison to
the morphological mix derived from the Stromlo-APM redshift survey (Loveday
et al. 1996), for galaxies brighter than $\magb=-18.8$ (E:S0:Sp = 13:11:76,
see \galone). However, such a disagreement is most probably not due to the
model itself, but rather to the different kind of environment considered in
the present paper (the immediate outskirts of galaxy clusters), and in the
Stromlo-APM redshift survey (the average field, which also includes clusters
and super-clusters, where the early-type fraction is higher). In fact, the
Stromlo-APM mix is better reproduced by our model when applied to large scale
cosmological simulations, where both high and low-density environments are
represented (see \galone).

\begin{figure*}
\centerline{\epsfxsize = 17.6 cm \epsfbox{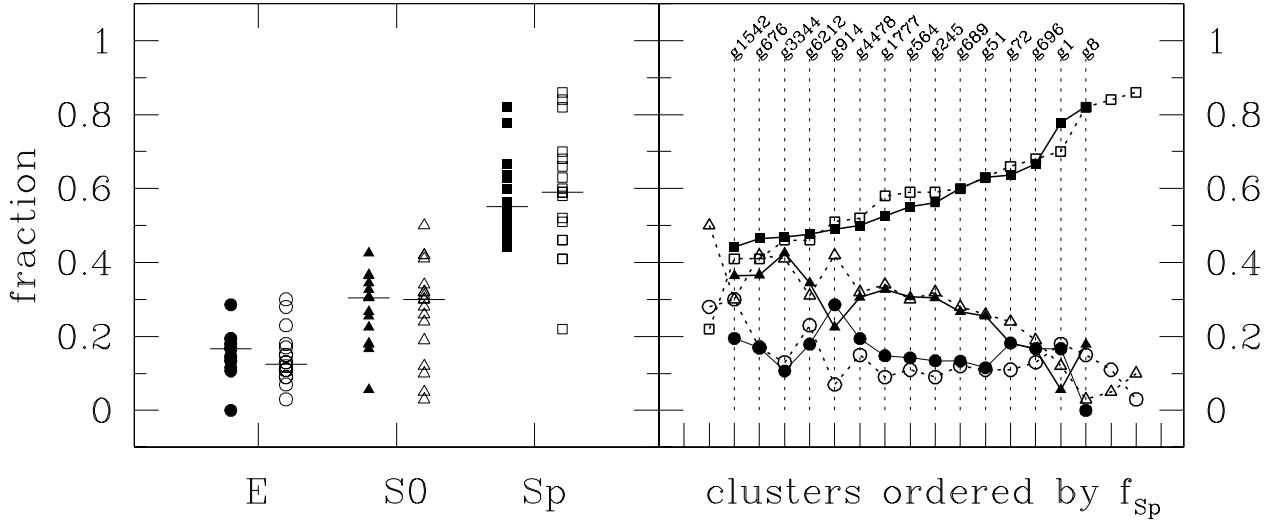}}
\caption{Comparison of the model morphological fractions to observations.
Model galaxies have been selected with $\magb<-18.8$ and within 2 Mpc$/h$
from cluster centres, similarly to the observations of Vogt et
al. (2004). {\it Left panel:} model results (solid symbols) are compared to
observational data (empty symbols) from Table 1 of Vogt et al. (2004). The
short horizontal lines mark the medians of the distributions. {\it Right
panel:} distribution of the relative morphological fractions (represented
with squares, triangles, and circles for spirals, lenticulars, and
ellipticals, respectively) for clusters ordered by increasing spiral
fraction, in the model (solid symbols), and in the observations (empty
symbols); clusters are sorted out along the x-axis by increasing spiral
fraction, with arbitrary spacing.}
\label{fig:vogt}
\end{figure*}

Moreover, \emph{the comparison with observations in clusters is very
satisfactory}, as apparent from Figure \ref{fig:vogt}. The data are from Vogt
et al. (2004), for all galaxies with measured position and redshift, and
within a distance of 2 Mpc$/h$ from the centres of a set of local
clusters. According to the authors, the sample ``starts to become
significantly incomplete at B band magnitudes ranging from $-18.5$ to
$-19.5$''. For a consistent comparison, we have therefore selected in our 15
simulations all galaxies with the same distance from cluster centres and with
$\magb<-18.8$.  The left panel of Figure~\ref{fig:vogt} indicates that the
distribution of morphological mixes found in the simulations is in good
agreement with that of the observed sample, with a large scatter from cluster
to cluster, and similar median values.  The right panel of
Figure~\ref{fig:vogt} shows that model results also reproduce the observed
anti-correlation between the global fraction of spirals and that of
lenticulars, with the global fraction of ellipticals staying almost constant
around 15\%.  Note that a radial distance 2 Mpc$/h$ is larger than the DM
halo virial radius for 10 of our 15 clusters, and it corresponds to at least
$0.7\times\rvir$ for the 5 most massive ones.  In Figure~\ref{fig:vogt}, we
are therefore considering the overall cluster population (brighter than
$\magb=-18.8$), not only the most central one, which explains why the spiral
fraction is comparable to or larger than that of early-type galaxies.
%

\subsection{Galaxy colours}
\label{sec:col}
It is well established that galaxy colours show a dependence on morphological
type, as well as on environment, with ellipticals and S0s being redder and
with tighter colour distributions than spirals, and with field galaxies
being, on average, bluer than their cluster counterparts.

As a preliminary remark, \emph{the colours of our simulated galaxies do not
appear to systematically depend on cluster richness}: the colour
distributions, both for early and late type galaxies, show approximately the
same shape in all clusters, independently of the mass of the host DM
halo. This statement holds for any colour and for any magnitude limit, as
well as for galaxies populating the cluster as a whole, or simply the
cores. Therefore, in the following, we will only focus on the {\it mean}
colour distributions, obtained by averaging over the entire 15 cluster
sample.

\begin{figure}
\centerline{\epsfxsize = 8.7 cm \epsfbox{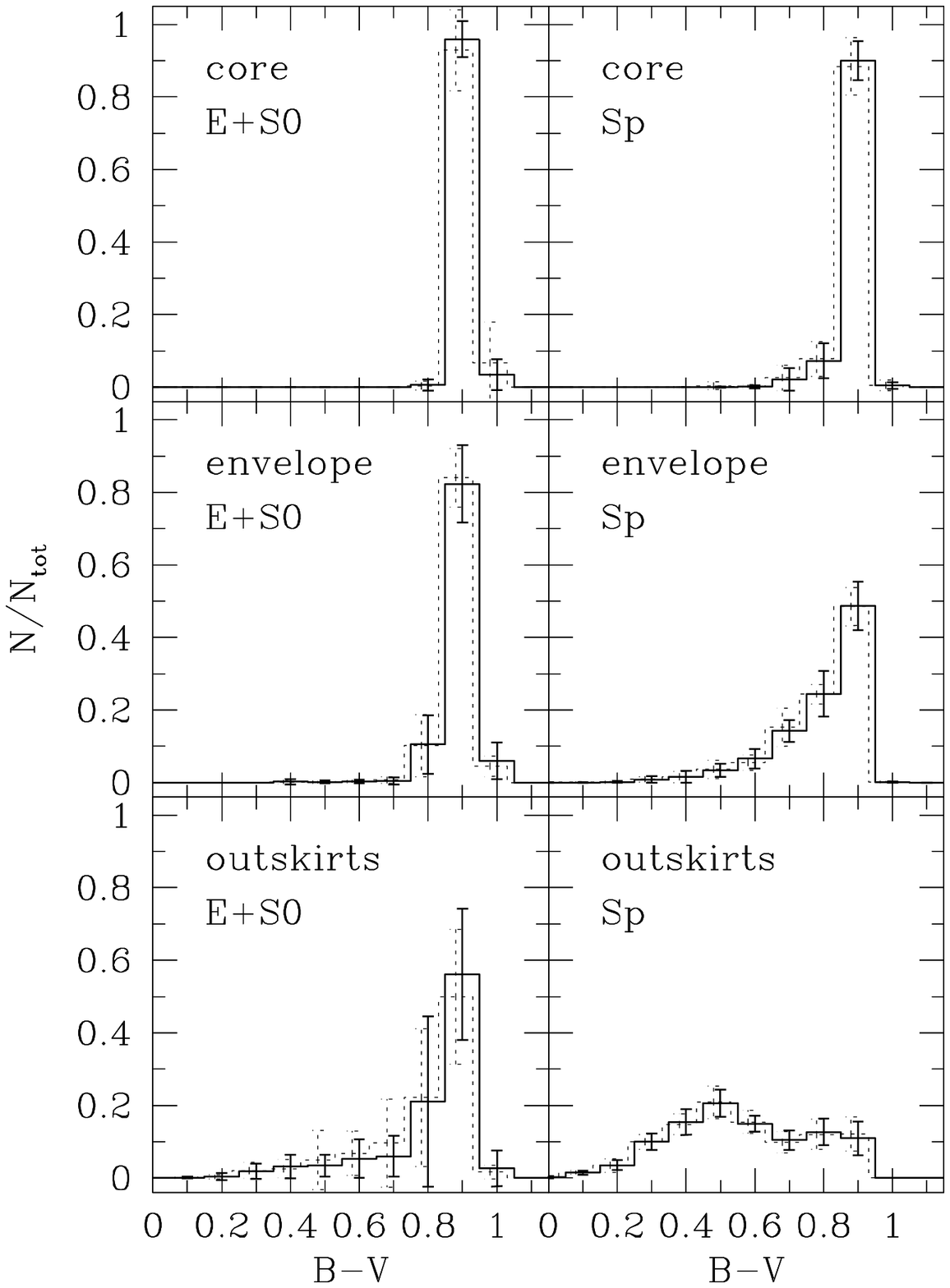}}
\caption{Dependence of galaxy B--V colours on environment.  Average B--V
colour distributions and rms dispersion about the mean, for early-type
galaxies ({\it left panels}) and spirals ({\it right panels}), in three types
of environment: cluster cores ({\it top panels}), envelopes ({\it middle
panels}) and outskirts ({\it bottom panels}). Dotted histograms show the
results if ram pressure stripping is not taken into account.}
\label{fig:BV_envir}
\end{figure}

We study the dependence of galaxy colour on environment in Figure
\ref{fig:BV_envir}, where the B--V distributions in cluster cores are
compared to those in the envelopes and in the outskirts, for early-type and
late-type galaxies, separately. For both morphological types, \emph{colours
appear to be extremely homogeneous in cluster cores, whereas a tail of bluer
galaxies develops when moving to regions of progressively lower density}. For
ellipticals and S0s, the peak of the distribution stays around B--V$\simeq
0.9$, and the scatter about it remains quite small in all environments.  The
colour distribution of spirals, instead, changes completely, from strongly
peaked around B--V$=0.9$ in cluster cores, to very dispersed around colours
more than 0.45 mag bluer, in the outskirts.  These trends do not change if we
restrict the analysis to sub-samples of brighter galaxies, or if we consider
other colours, and are in qualitative agreement with the observations.
%


\begin{figure*}
\centerline{\epsfxsize = 17.6 cm \epsfbox{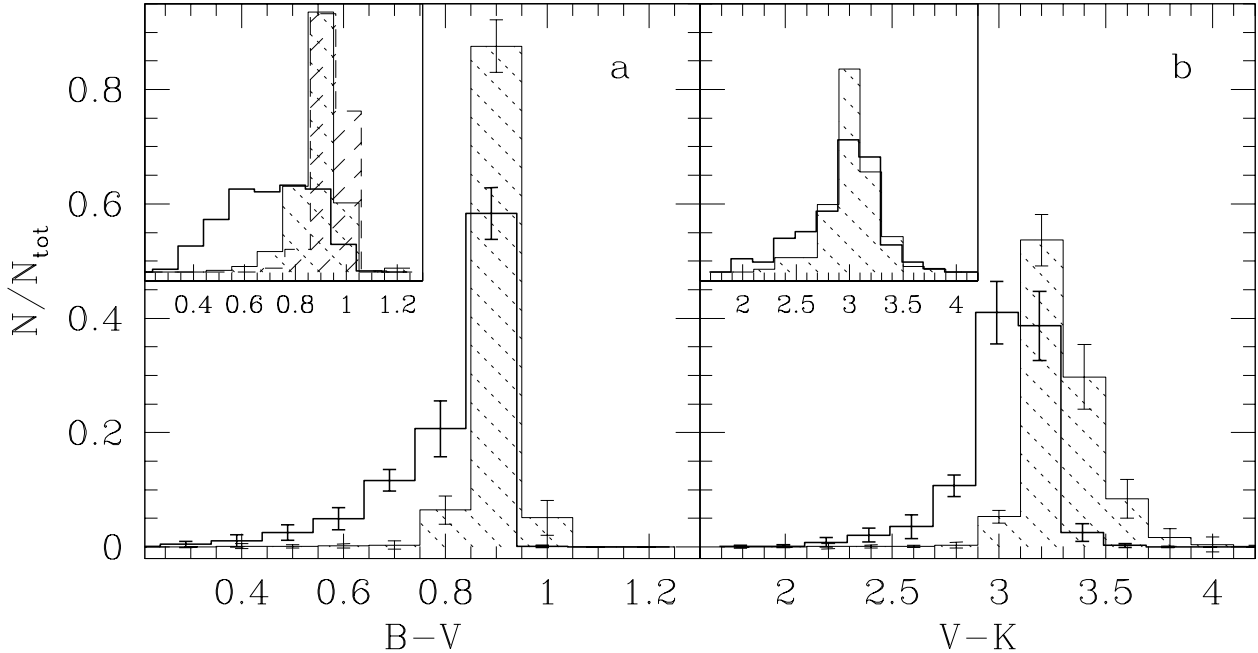}}
\caption{Dependence of galaxy B--V and V--K colours on morphological
type. {\it Panel a:} average B--V colour distribution of cluster ellipticals
and S0s (shaded histograms), compared to that of spirals (empty
histogram). Only model galaxies within $\rvir$ and with $\magb<-17$ are
considered. The analogous histograms in the insert (where values along the
y-axis are arbitrary) refer to observed galaxies obtained from GOLDMine. The
additional histogram in the left panel insert (shaded in long-dashed line)
represents early-type galaxies from the Prugniel \& Simien (1996)
sample. {\it Panel b:} the same as in panel $a$, but for the V--K colour, and
for galaxies brighter than $\magv =-18.2$.  The insert shows the observed
distribution obtained from GOLDMine.  All observations are extinction and
k-corrected (see text).}
\label{fig:ES0_BV_VK}
\end{figure*}

A more quantitative study of the dependence of colours on morphological types
is provided in Figure \ref{fig:ES0_BV_VK}, where we directly compare the B--V
and the V--K colour distributions of cluster ellipticals and S0s (dashed
histograms), to those of spirals (empty histograms).  For both colours,
\emph{spirals appear to be systematically bluer than early-type galaxies}. In
B--V (panel a), the vast majority of early-type galaxies is well
characterised by a single colour (B--V$\simeq 0.9$) with a small (0.1 mag)
dispersion, while spirals redder than B--V$=0.9$ are not found, and their
distribution is wider (more dispersed) and shifted towards bluer colours.  In
V--K, the distributions are more dispersed and skewed towards red and blue
colors for early and late type galaxies, respectively (panel b).
Observational results are shown in the inserts of Figure \ref{fig:ES0_BV_VK},
where data are taken from the {\em Galaxy On Line Database Milano Network}
(hereafter GOLDMine, Gavazzi et al. 2003), which we corrected for galactic
extinction, as estimated by Schlegel et al. (1998), using the NASA/IPAC
Extragalactic Database (NED), and k-corrected following Poggianti (1997). For
the B--V colours we also show data from Prugniel \& Simien (1996), which were
already extinction and k-corrected.  The general trends found for our model
results are in qualitative agreement with the observations. More in detail,
\emph{the V-K distribution of simulated spirals well corresponds to the
observed one, while their B-V colours are not blue enough and present and
unobserved peak around B--V$\simeq 0.9$}.  This is a drawback of the model,
possibly due to the fact that gas accretion is not allowed on satellite
galaxies in any DM halo, so that fewer new stars can form as redshift
decreases, thus leading to older and redder stellar populations than the
observed ones.  \emph{In B--V, a much better agreement is found for
ellipticals and S0s}, with the right colour for the peak of the distribution,
and only a relatively small deficiency of slightly bluer (B--V$\simeq 0.8$)
and slightly redder (B--V$\simeq 1$) galaxies.  However, \emph{in V--K the
whole colour distribution of ellipticals and S0s is 0.2 mag too red with
respect to the observations}.

\begin{figure*}
\centerline{\epsfxsize = 17.6 cm \epsfbox{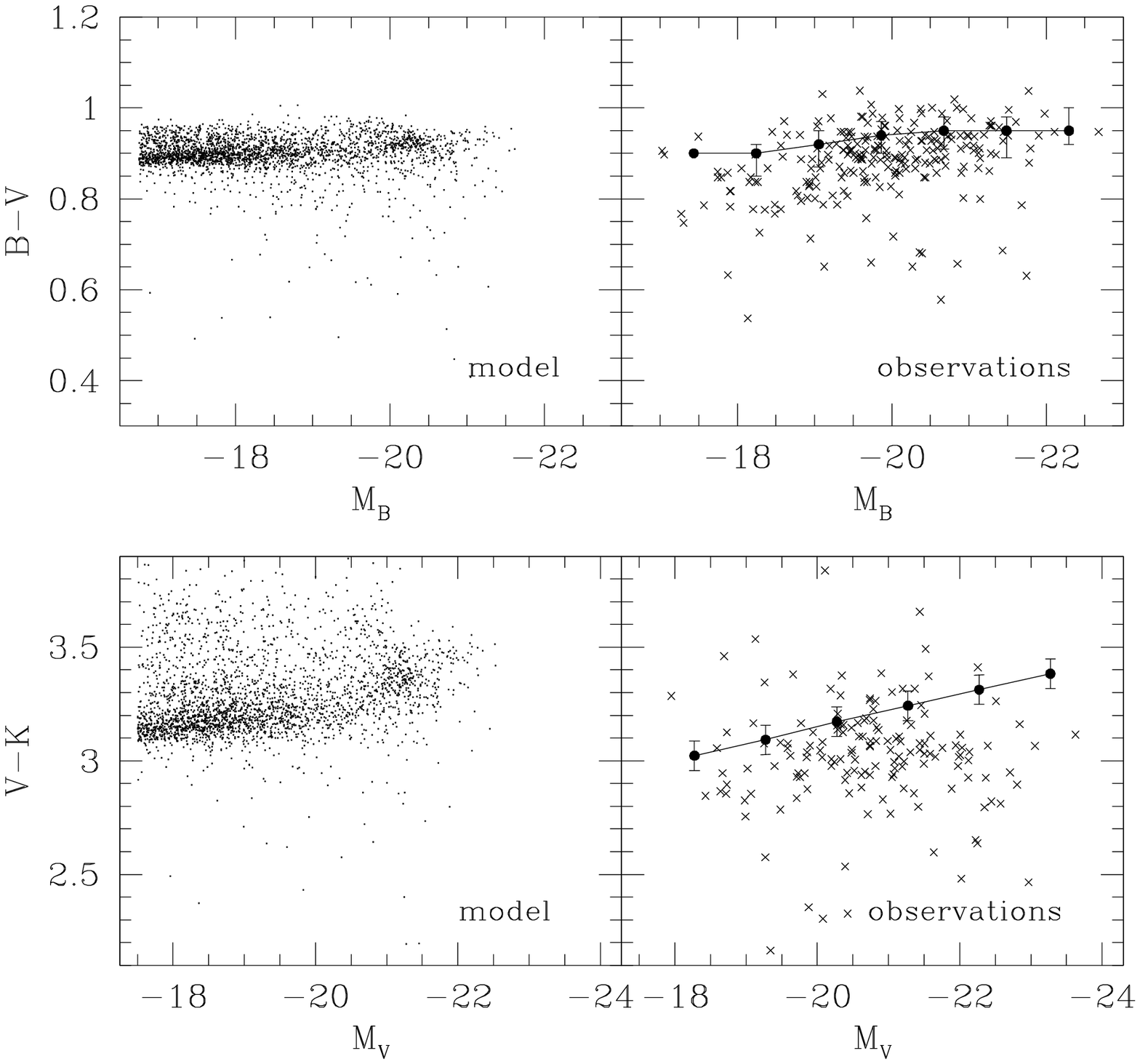}}
\caption{Colour-magnitude relations for cluster early-type galaxies. Model
results are shown in the {\it left-hand panels}, observations (extinction and
k-corrected) in the {\it right-hand panels}. {\it Upper panels:} B--V colour
as a function of the B absolute magnitude. Model results are shown for all
early-type galaxies within the virial radius of the 15 simulated clusters.
Observational data extracted from the GOLDMine data base are shown as
crosses, while the median colour-magnitude relation, with rms dispersion
error bars, obtained from the data of Prugniel \& Simien (1996) is plotted as
a solid line.  {\it Lower panels:} the same as in the upper panels, but for
the V--K colour as a function of the V magnitude. Observations refer to the
GOLDMine data (crosses), and to the mean colour-magnitude relation, with
standard deviation error bars, for early-type galaxies in Coma cluster, from
Bower et al. (1992; solid line).}
\label{fig:CM}
\end{figure*}

The model predictions for early-type galaxies are further compared to
observations in Figure \ref{fig:CM}, in terms of the colour-magnitude
relations (data are as in Figure~\ref{fig:ES0_BV_VK}, plus those from Bower,
Lucey \& Ellis 1992 in V-K. To the latter, which are already extinction
corrected, we applied Poggianti's k-corrections).  Simulated galaxies do
follow tight and well defined relations, especially for the B--V colour.  In
this case, both the slope and the scatter in model results reproduce well the
observed ones.  The V--K colour-magnitude relation less accurately matches
the observations, showing an average shift of $\sim 0.2$ mag, a too large
scatter (too skewed towards red colours), and a flatter slope.  As already
discussed in Section \ref{sec:LF}, a discrepancy also exists in terms of a
lack in the model of the brightest cluster galaxies ($\magb\lsim-22$ and
$\magv\lsim -22.5$) found in the observations.
%

Note however that the exact shape of the colour distribution depends quite
sensitively on the definition of morphological types, which is not exactly
the same in the model and the observations. Moreover, the way colours have
been obtained (from the galaxy total light, as in the model, or within
smaller central regions due to fixed aperture observations, for instance), as
well as the adopted extinction and k- corrections, can easily lead to
discordant results and therefore make it difficult to compare observations
and models (see e.g. Kaviraj et al. 2004).

\subsection{Effects of ram pressure stripping}
We find that \emph{by including or neglecting ram pressure stripping in the
model, galaxy properties only show mild variations}. The LFs obtained by
taking into account or by ``switching off'' this process are almost
indistinguishable. This is shown, for instance, in Figure \ref{fig:lf_env}
for cluster core environment, where the stripping should be maximally
efficient.  Also the effects on galaxy color distributions are completely
negligible, as can be seen in Figure \ref{fig:BV_envir}.  The morphological
mix in cluster cores appears to be mildly affected by ram pressure stripping,
with a slight increase of the fraction of ellipticals and spirals, in spite
of that of lenticulars, if the process is neglected (panel a in Figure
\ref{fig:morph_env}). However, the effect is small and no systematic trends
with cluster richness are found.  The cold gas content of spiral galaxies is
effectively smaller in clusters than in the outskirts, but this is mainly due
to the different gas accretion, gas consumption, and star formation histories
in the two kinds of environment: in dense environments, objects form earlier
on average and therefore have shorter dynamical and star formation
timescales.  Moreover, field spirals in the model typically correspond to the
central galaxy of the embedding DM haloes, and thus they can continuously
accrete new cold gas onto their disks; cluster spirals, instead, are
satellite galaxies, which can only consume their gas reservoir until its
complete exhaustion.  Thus, by the time the ram pressure stripping process
becomes efficient (cluster assembly epoch), most of the gas has already been
processed in a large fraction of the cluster member galaxies, and similar
results are found whether or not we include ram pressure in the model.  We
therefore confirm the conclusions of Okamoto \& Nagashima (2003) that ram
pressure stripping plays a minor role in galaxy transformations, provided a
more accurate description of this process does not prove it to be more
efficient by, say, an order of magnitude than our estimates.

\section{Summary and discussion}
\label{sec:discuss}
In this paper, we have studied galaxy formation and evolution in clusters and
in their outskirts, within the framework of the hierarchical merging
scenario, by means of high-resolution re-simulations of massive DM haloes,
and the \gal\ hybrid galaxy formation model.

Numerical N-body simulations and a good mass resolution are both necessary to
accurately follow the formation history of DM haloes, where observable
galaxies form and evolve. An effective method, in terms of computational
cost, consists in re-simulating with a larger number of particles a given
halo or a given region selected from an already-run N-body simulation, thus
increasing its mass resolution.  In the present work, we have applied such a
technique to a sample of 15 massive DM haloes, increasing their mass
resolution by a factor $\sim 33$ or 66, so that the evolution of all DM
haloes of at least 2 or $1\times 10^{10}\msol/h$ can now be followed.

To detect haloes, construct their merging histories, and form galaxies within
them, we have employed the \gal\ model, that has been presented and tested
against several observations of the local and high redshift Universe in the
first and third papers of this series (Hatton et al. 2003; Blaizot et
al. 2004).  Here, \gal\ is used to investigate whether (and how) galaxy
properties depend on cluster richness (with our halo masses in the range
between $10^{14}$ and $10^{15}\msol$), and on the density of the local
environment (cluster cores, envelopes and outskirts).  With respect to
\galone, the improved mass resolution of our simulations have raised the
necessity for a better description of the reionisation history of the
Universe, needed to avoid an unobserved excess of galaxies at the faint-end
of the LF. A semi-analytic description of ram pressure stripping has also
been included, since this process is commonly considered to be important in
cluster environment.  Luminosity functions, morphological fractions and
colours have been studied for all galaxies brighter than the magnitude limits
imposed by our mass resolution: $\magb=-16.75$, $\magv=-17.5$, $\magk=
-20.5$, in B, V and K bands, respectively.  We now proceed to summarise and
discuss our main results.


\subsection{Luminosity function}
\begin{itemize}
\item[(i)] The cluster LF appears to be universal, in all photometric bands:
its shape varies from cluster to cluster, but it does not show any clear
dependence on cluster richness (mass), at least in the galaxy luminosity and
cluster mass ranges probed by the model.
\item[(ii)] A gradual steepening of the faint-end of the LF with increasing
density of the local environment (i.e., from cluster outskirts, to cluster
envelopes, to the cores) is detected in all bands.
\item[(iii)] In K, the brightest galaxies have similar luminosities both in
cluster cores and in the outskirts, while in B, cluster central galaxies
never attain the same high luminosities as their outskirts counterparts.
\item[(iv)] The LF of cluster early-type galaxies is flatter than that of
spirals.
\item[(v)] A good agreement is found with respect to the observed cluster
LFs, both in B and in K, and also for different morphological types
considered separately.  The main discrepancy consists in too faint
luminosities of the brightest cluster members: galaxies brighter than
$\magb\simeq -22$ and $\magk\simeq -26$ are absent in the model, but are
found in the observations.
\item[(vi)] The agreement is still satisfactory between the model LFs for the
cluster outskirts and the observed LFs for the average field in the K
band. In B, instead, the model predicts a too bright $\mstar$, and a too flat
faint-end slope, compared to the observed values.
\end{itemize}

Some notes of caution should be spent about the comparison between
simulations and observations.  For what concerns the model LF, its faint-end
is very sensitive to the magnitude resolution limits and to the detailed
modelling of reionisation squelching at high redshift, a process that is
still poorly known and that also depends, in turn, on the mass resolution of
the DM simulations (see Section \ref{sec:new_ele}). The bright-end, instead,
is affected by galaxy merging, with both the frequency of this process and
the exact characteristics of the merger end-products being modelled in a very
approximated way.  Finally, the exact values of the Schechter best fit
parameters are not very robust, since they depend on the considered binning
and range of magnitudes.  On the side of observations, a general consensus on
the exact shape of the LF, both in clusters and in the field, and on its
dependence on the local density still lacks (see, e.g., Beijersbergen et
al. 2002, Liske et al. 2003, Lin et al. 2004, and references
therein). Results are quite sensitive to several factors, like the considered
range of magnitudes, the inclusion/exclusion of the brightest cluster
members, the adopted corrections (for incompleteness of the sample, surface
brightness effects, etc.), and the method used for combining different
samples and get the average LF. Also the way fits are performed can affect
the results, with a single Schechter function often providing a poor
description of the observed LFs (e.g., Popesso et al. 2004, and references
therein). A further difficulty for a proper comparison between model and
observations, as well as among different observational results, is added by
the necessary transformation from one photometric band to another, that pass
through uncertain and approximated average colour conversions.

We finally stress that our comparisons are mainly valid in terms of the
overall shape of the LFs, while they do not provide information about the
relative excess or deficiency of galaxies, since they depend on the adopted
normalisation (as it is always the case when comparing functional shapes that
are not identical): for instance, if we normalise our outskirts LF to the
number of galaxies more luminous than $\magb=-20.5$ (rather than
$\magb=-18.5$), we better recover the bright-end of the observed field LF,
but further underestimate the fraction of faint objects.  If one also
considers that the types of environment, as well as the definitions of
morphological types do not exactly correspond in the model and in the
observational studies, the agreements we find should be considered quite
satisfactory. 

\subsection{Morphological fractions}
\begin{itemize}
\item[(i)] No clear trend is found between the relative fractions of spirals,
S0s and ellipticals and cluster richness in our results.
\item[(ii)] The morphological mix shows quite a large scatter from cluster to
cluster, and a significant dependence on the adopted magnitude cut,
particularly in cluster cores. In fact, whereas for $\magb<-17$ spirals are
more abundant than ellipticals in the cores of all clusters, early-type
galaxies dominate in most of the cases for $\magb<-19$.
\item[(iii)] The morphological mix clearly changes with environment, from
massively spiral-dominated in the outskirts, to a more balanced mix between
spirals and ellipticals/S0s when moving to cluster cores.
\item[(iv)] A good agreement with the observations is found for the
morphological mix in clusters.
\item[(v)] With respect to the observational data for the ``average field'',
we find an excess of spirals (and a deficiency of early-type galaxies) in our
cluster outskirts.
\end{itemize}

While this latter discrepancy is probably related to the different kind of
environment considered in the model and in the observations, the success at
reproducing the cluster morphological mix is not trivial, particularly if one
considers that different definitions of morphological types are adopted.
Since galaxy positioning in the model is currently described in a rough way,
we do not push the comparison further and try to reproduce the observed
cluster population gradients in more details, but we postpone this issue to a
forthcoming paper.

\subsection{Colour distributions}
\begin{itemize}
\item[(i)] Galaxy colour distributions do not appear to depend on cluster
richness, whatever our adopted magnitude limits, or whatever sub-samples of
galaxies we select in different environments.
\item[(ii)] Early-type galaxies are systematically redder than spirals in all
environments.
\item[(iii)] Colours are very uniform in cluster cores, with a well defined peak
around B--V=0.9, both for early and late type galaxies.
\item[(iv)] The distributions become increasingly dispersed towards bluer
colours when moving to environment of lower densities, still maintaining a
peak at B--V$\simeq 0.9$ in the case of the ellipticals and S0s, while
completely changing to a flat and bluer distribution for spirals in cluster
outskirts.
\item[(v)] Compared to observational data from a sample of local clusters,
the B--V colours of simulated spirals appear to be too uniform and too red, while
a better agreement is found for the V--K. 

\item[(vi)] The B--V colour distribution and colour--magnitude relation of
early-type galaxies in clusters is in fairly good agreement with the observed
ones, whereas a too flat slope, an average shift of $\sim 0.2$ mag, and too
large dispersion towards the red is found for the V--K colours.
\end{itemize}

The observed general trends of galaxy colours with morphological type (redder
for early-type galaxies) and environment (redder and more uniform in denser
environments) are thus reproduced by our simulations, but some discrepancies
are also apparent.  Disagreements are in the sense of not-blue-enough or
too-red colours, and might be due to an imprecise dust modelling, as well as
to the fact that gas accretion, and the consequent formation of new stars, is
not allowed on non-central galaxies. Moreover, different factors contribute
to make difficult an appropriate comparison between observations and model
results: not only the already discussed differences in the adopted
definitions of morphological types and environments, but also the way colours
have been obtained and, in certain cases, corrected.  In fact, the known
internal colour gradients in galaxies, imply that data obtained from the
galactic total light (as in our simulations), or from the luminosity
integrated on smaller, central areas (possibly corrected afterwards for
aperture effects) may substantially differ and bring to discordant
colour--magnitude relations (e.g., Bender, Burstein \& Faber 1993; Fioc \&
Rocca-Volmerange 1999; Scodeggio 2001; Kaviraj et al. 2004).  When feasible,
we have tried here to homogeneize the model and observational results as much
as possible, but part of the discrepancies might still be ascribable to
aperture effects and errors in extinction and k-corrections.

\subsection{Ram pressure stripping}
By means of a simplified formula, we have modelled the ram pressure stripping
of cold gas from galactic discs, during the motion of galaxies through the
diffuse hot intra-cluster medium.  All results appear to be practically
unaffected by the inclusion of this process in the model, in agreement with
the conclusions of Okamoto \& Nagashima (2003). We stress, however, that a
more accurate modelling of this process is required before completely
assessing its role in modifying galaxy properties during their cosmic
evolution.

\subsection{Conclusions}
The present paper is part of the series investigating the ability of the
\gal\ model to reproduce and predict a full set of observational data, in
several photometric bands and environments, and at various redshifts.  Here
we have focussed on clusters of galaxies of different richnesses, at $z=0$.
A general good agreement between observations and theoretical results is
found, even if some drawbacks are also apparent in the model. The main
problems consist in a too faint luminosity of the brightest cluster galaxies,
and a not yet satisfactory reproduction of galaxy colours, which tend to be
too red (or not blue enough).  While the present analysis concerns a limited
set of galaxy properties in local clusters, we plan to extend it to other
observables and to higher redshifts, thus addressing important issues, like,
for instance, the Butcher--Oemler effect (Butcher \& Oemler 1984), and the
origin and time evolution of galaxy scaling relations, all of which should
further constrain the model.  Moreover, some improvement to \gal\ is also
planned, and is expected to help to better reproduce the observations. In
particular, we intend to describe galaxy positioning and interaction rates
within DM haloes in a more accurate way, i.e., by directly following
sub-structures in the DM N-body simulations and associating the baryonic
galaxies to them.  This is expected to have several and relevant effects on
final results (e.g., Springel et al. 2001b), since many galaxy properties
(like the morphology, the stellar content, the overall luminosity and the
dynamical characteristics) depend more or less strongly on the merging
history of the galaxy.  Allowing gas cooling on satellite (non central)
galaxies, taking into account other sources of feedback, like type Ia
supernovae, and following the growth of central super massive black holes and
their effects on galaxy properties should also help to describe galaxy
formation in a more suitable way.  Then, all the analysis in the papers of
this series, already published and in preparation, will finally give a
complete judgement of the ability of the \gal\ model, and, in turn, of the
hierarchical merging scenario, to describe how galaxies form and evolve in
the Universe.

\section*{Acknowledgments}
B.L. thanks Jeremy Blaizot, Lucia Pozzetti, Gianni Zamorani, Luca Ciotti and
Simon White for many useful discussions.  B.L. is also in debt with Volker
Springel, Bepi Tormen, and Naoki Yoshida for their help with the N-body
simulations. We acknowledge the anonymous referee for useful comments.
Simulations have been run with GADGET (Springel et al. 2001a) on the IBM SP2
of the CINES (Montpellier, France), and the CRAY T3E of the RZG Computing
Center (Munich, Germany). This research has made use of the GOLDMine Database
(Gavazzi et al. 2003), and the NASA/IPAC Extragalactic Database (NED) which
is operated by the Jet Propulsion Laboratory, California Institute of
Technology, under contract with the National Aeronautics and Space
Administration.  B.L. is supported by a post-doc fellowship by Italian INAF.

\bsp
\label{lastpage}

\end{document}